\documentclass[12pt,twoside,english,draftclsnofoot,onecolumn]{IEEEtran}
\usepackage[T1]{fontenc}
\usepackage[latin9]{inputenc}
\usepackage{amsmath}
\usepackage{amsthm}
\usepackage{amssymb}
\usepackage{graphicx}
\PassOptionsToPackage{normalem}{ulem}
\usepackage{ulem}
\usepackage{cite}
\makeatletter

\providecommand{\tabularnewline}{\\}

\theoremstyle{plain}
\newtheorem{thm}{\protect\theoremname}
\theoremstyle{definition}
\newtheorem{defn}[thm]{\protect\definitionname}
\theoremstyle{plain}
\newtheorem{cor}[thm]{\protect\corollaryname}
\theoremstyle{plain}
\newtheorem{prop}[thm]{\protect\propositionname}

\makeatother

\usepackage{babel}
\providecommand{\corollaryname}{Corollary}
\providecommand{\definitionname}{Definition}
\providecommand{\propositionname}{Proposition}
\providecommand{\theoremname}{Theorem}

\begin{document}

\title{Polar Codes and Polar Lattices for the Heegard-Berger Problem}

\author{\IEEEauthorblockN{Jinwen Shi, Ling Liu, Deniz Gündüz, \emph{Senior Member, IEEE}, \\
and Cong Ling, \emph{Member, IEEE}}}
\maketitle
\begin{abstract}
Explicit coding schemes are proposed to achieve the rate-distortion
function of the Heegard-Berger problem using polar codes. Specifically,
a nested polar code construction is employed to achieve the rate-distortion
function for the doubly-symmetric binary sources when the side information
may be absent. The nested structure contains two optimal polar codes
for lossy source coding and channel coding, respectively. Moreover,
a similar nested polar lattice construction is employed when the source
and the side information are jointly Gaussian. The proposed polar
lattice is constructed by nesting a quantization polar lattice and
a capacity-achieving polar lattice for the additive white Gaussian
noise channel.
\end{abstract}

\begin{IEEEkeywords}
Heegard-Berger Problem, source coding, lattices.
\end{IEEEkeywords}

\let\thefootnote\relax

\footnote{Jinwen Shi, Ling Liu, Deniz Gündüz and Cong Ling are with Dept. of
Electronic and Electrical Engineering, Imperial College London, SW7
2AZ, UK. (e-mails: \{jinwen.shi12, l.liu12, d.gunduz\}@imperial.ac.uk,
cling@ieee.org).} 


\section{Introduction}

\begin{figure}[t]
\begin{centering}
\includegraphics[scale=0.65]{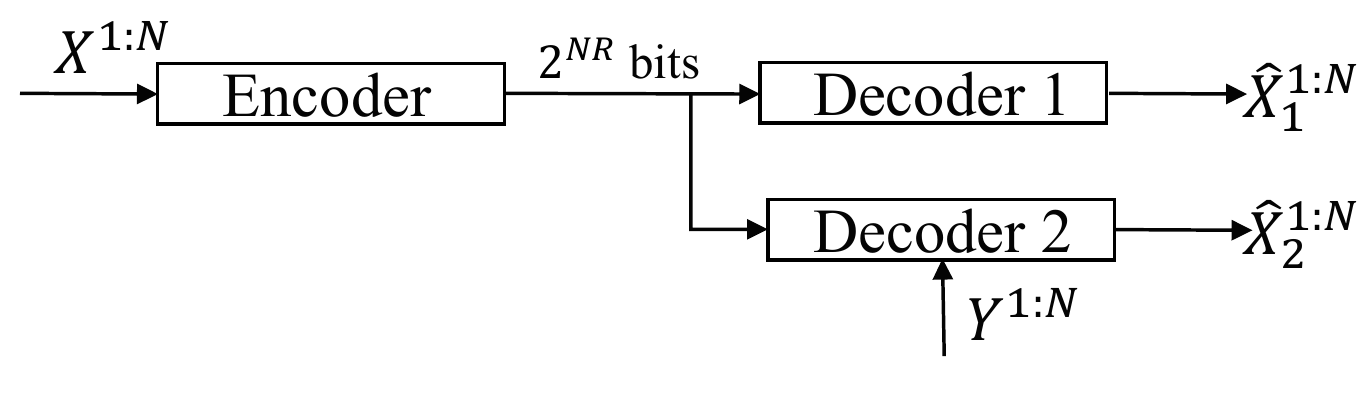}
\par\end{centering}
\vspace{-0.5cm}

\caption{Illustration of the Heegard-Berger rate-distortion problem.\label{fig:HB-codingsystem} }

\vspace{-0.3cm}
\end{figure}

The well-known Wyner-Ziv problem is a lossy source coding problem
in which a source sequence is to be reconstructed in the presence
of correlated side information at the decoder \cite{Wyner1976Ziv}.
An interesting question is whether reconstruction with a non-trivial
distortion quality can still be obtained in the absence of the side
information. The equivalent coding system contains two decoders, one
with the side information, and the other without, as shown by Fig.
\ref{fig:HB-codingsystem}. 

In 1985, Heegard and Berger \cite{Heegard1985absent} characterized
the rate-distortion function $R_{HB}\left(D_{1},D_{2}\right)$ for
this scenario, where $D_{1}$ is the distortion achieved without side
information, $D_{2}$ is the distortion achieved with it, and $R_{HB}\left(D_{1},D_{2}\right)$
denotes the minimum rate required to achieve the distortion pair $\left(D_{1},D_{2}\right)$.
They also gave an explicit expression for the quadratic Gaussian case.
Kerpez \cite{Kerpez1987HB} provided upper and lower bounds on the
Heegard-Berger rate-distortion function (HBRDF) for the binary case.
Later, the explicit expression for $R_{HB}\left(D_{1},D_{2}\right)$
in the binary case was derived in \cite{Tian2006HB} together with
the corresponding optimal test channel. Our goal in this paper is
to propose explicit coding schemes that can achieve the HBRDF for
binary and Gaussian distributions. 

The Heegard-Berger problem is a generalization of the classical Wyner-Ziv
problem, in which a source sequence is to be reproduced at the decoder
within a certain distortion target, and the side information available
at the decoder is not available at the encoder. A nested construction
of polar codes is presented in \cite{polarchannelandsource} to achieve
the binary Wyner-Ziv rate-distortion function. For Gaussian sources,
a polar lattice to achieve both the standard and Wyner-Ziv rate-distortion
functions is proposed in \cite{PolarlatticeQZ}. Different from the
solutions of the Wyner-Ziv problem in \cite{polarchannelandsource}
and \cite{PolarlatticeQZ}, we need to consider the requirements for
the two decoders jointly. Therefore, we make use of the low-fidelity
reconstruction at Decoder 1, and combine it with the original source
and the side information to form the nested structure that achieves
the optimal distortion for Decoder 2. 

The optimality of polar codes for the lossy compression of nonuniform
sources is shown in \cite{aspolarcodes}. We employ this scheme as
part of our solution, since the optimal forward test channel may be
asymmetric in the binary Heegard-Berger problem. Furthermore, it is
shown in \cite{Ye2015Barg} that polar codes are optimal for general
distributed hierarchical source coding problems. The Heegard-Berger
problem can also be considered as a successive refinement problem.
In this paper, we propose explicit coding schemes using polar codes
and polar lattices to achieve the theoretical performance bound in
the Heegard-Berger problem. Practical codes for the Gaussian Heegard-Berger
problem are also developed in \cite{HB2011TrellisLDPC} which hybridize
trellis and low-density parity-check codes. However, the optimality
of this scheme to achieve the HBRDF is not shown in \cite{HB2011TrellisLDPC}.

The contributions of this paper can be summarized as follows:
\begin{itemize}
\item We propose a nested construction of polar codes for the non-degenerate
region of the binary Heegard-Berger problem, and prove that they achieve
the HBRDF for doubly symmetric binary sources (DSBS). We consider
the reconstruction of the source sequence at Decoder 1, i.e., the
decoder without side information, denoted by $\hat{X}_{1}^{1:N}$
and the original source sequence $X^{1:N}$ as a combined source,
and further combine this reconstruction $\hat{X}_{1}^{1:N}$ with
the original side information $Y^{1:N}$ to obtain a combined side
information. By this argument, we obtain another nested construction
of polar codes, which achieves the HBRDF of the entire non-degenerate
region. In addition, we present an explicit coding scheme by using
two-level polar codes to achieve the HBRDF whose forward test channel
may be asymmetric. Finally, we prove that polar codes achieve an exponentially
decaying block error probability and excess distortion at both decoders
for the binary Heegard-Berger problem.
\item We then consider the Gaussian Heegard-Berger problem, and propose
a polar lattice construction that consists of two nested polar lattices,
one of which is additive white Gaussian noise (AWGN) capacity-achieving
while the other is Gaussian rate-distortion function achieving. This
construction is similar to the one proposed for the Gaussian Wyner-Ziv
problem in \cite{PolarlatticeQZ}. However, in the Heegard-Berger
problem setting, we need to treat the difference between the original
source and its reconstruction at Decoder 1 as a new source, and the
difference between the original side information and the reconstruction
at Decoder 1 as a new side information. As a result, we can obtain
an optimal test channel that connects the new source with the new
side information by using additive Gaussian noises. According to this
test channel, we can further construct two nested polar lattices that
achieve the Gaussian HBRDF of the entire non-degenerate region.
\end{itemize}

\textit{Organization:} The paper is organized as follows: Section
II presents the background on binary and Gaussian Heegard-Berger problems.
The construction of polar codes to achieve the HBRDF for DSBS is investigated
in Section III. In Section IV, the polar code construction for the
Gaussian Heegard-Berger problem is addressed. The paper is concluded
in Section V.

\textit{Notation:} All random variables are denoted by capital letters,
while sets are denoted by capital letters in calligraphic font. $P_{X}$
denotes the probability distribution of a random variable $X$ taking
values in set $\mathcal{X}$. For two positive integers $i<j$, $x^{i:j}$
denotes the vector $\left(x^{i},\ldots,x^{j}\right)$, which represents
the realizations of random variables $X^{i:j}$. For a set $\mathcal{F}$
of positive integers, $x_{\mathcal{F}}$ denotes the subvector $\{x^{i}\}_{i\in\mathcal{F}}$.
For the Gaussian case, we construct polar codes in multiple levels,
in which $X_{l}$ denotes a random variable at level $l$, and $x_{l}^{i}$
its $i$-th realization. Then, $x_{l}^{i:j}$ denotes the vector $\left(x_{l}^{i},\ldots,x_{l}^{j}\right)$,
and $x_{l}^{\mathcal{F}}$ denotes the subvector $\{x_{l}^{i}\}_{i\in\mathcal{F}}$
at the $l$-th level. $\mathcal{F}^{c}$ and $\left|\mathcal{F}\right|$
denote the complement and cardinality of set $\mathcal{F}$, respectively.
For a positive integer $N$, we define $\left[N\right]\triangleq\left\{ 1,\ldots,N\right\} $.
$\boldsymbol{1}\left[x\in\mathcal{X}\right]$ denotes the indicator
function, which equals $1$ if $x\in\mathcal{X}$ and $0$ otherwise.
Let $I\left(X;Y\right)$ denote the mutual information between $X$
and $Y$. In this paper, all logarithms are base two, and information
is measured in bits.

\section{Problem Statement}

\subsection{Heegard-Berger Problem}

Let $\left(\mathcal{X},\mathcal{Y},P_{XY}\right)$ be discrete memoryless
sources (DMSs) characterized by the random variables $X$ and $Y$
with a generic joint distribution $P_{XY}$ over the finite alphabets
$\mathcal{X}$ and $\mathcal{Y}$. 
\begin{defn}
An $\left(N,M,D_{1},D_{2}\right)$ Heegard-Berger code for source
$X$ with side information $Y$ consists of an encoder $f^{(N)}:\mathcal{X}^{1:N}\rightarrow\left[I_{M}\right]$
and two decoders $g_{1}^{(N)}:\left[I_{M}\right]\rightarrow\mathcal{\hat{X}}_{1}^{1:N}$;
$g_{2}^{(N)}:\left[I_{M}\right]\times\mathcal{Y}^{1:N}\rightarrow\mathcal{\hat{X}}_{2}^{1:N}$,
where $\mathcal{\hat{X}}_{1},\mathcal{\hat{X}}_{2}$ are finite reconstruction
alphabets, such that 
\[
\begin{aligned}\mathbb{E}\left[\frac{1}{N}\sum_{j=1}^{N}d\left(X^{j},\hat{X}_{i}^{j}\right)\right] & \leq D_{i},\textrm{ }i=1,2,\end{aligned}
\]
where $\mathbb{E}$ is the expectation operation, and $d\left(\cdot,\cdot\right)<\infty$
is a per-letter distortion measure. In this paper, we set $d\left(\cdot,\cdot\right)$
to be the Hamming distortion for binary sources, and the squared error
distortion for Gaussian sources. 
\end{defn}

\begin{defn}
Rate $R$ is said to be $\left\{ \left(D_{1},D_{2}\right)-achievable\right\} $,
if for every $\epsilon>0$ and sufficiently large $N$ there exists
an $\left(N,M,D_{1}+\epsilon,D_{2}+\epsilon\right)$ code with $R+\epsilon\geq\frac{1}{N}\log M$.
\end{defn}
The HBRDF, $R_{HB}\left(D_{1},D_{2}\right)$, is defined as the infimum
of $\left(D_{1},D_{2}\right)$-achievable rates. A single-letter expression
for $R_{HB}\left(D_{1},D_{2}\right)$ is given in the following theorem.
\begin{thm}
(\cite[Theorem 1]{Heegard1985absent})
\[
R_{HB}\left(D_{1},D_{2}\right)=\hspace{-6bp}\hspace{-10bp}\min_{\left(U_{1},U_{2}\right)\in\mathcal{P}\left(D_{1},D_{2}\right)}\hspace{-10bp}\left[I\left(X;U_{1}\right)+I\left(X;U_{2}|U_{1},Y\right)\right],
\]
where $\mathcal{P}\left(D_{1},D_{2}\right)$ is the set of all auxiliary
random variables $\left(U_{1},U_{2}\right)\in\mathcal{U}_{1}\times\mathcal{U}_{2}$
jointly distributed with the generic random variables $\left(X,Y\right)$,
such that: i) $Y\leftrightarrow X\leftrightarrow\left(U_{1},U_{2}\right)$
form a Markov chain; ii) $\left|\mathcal{U}_{1}\right|\leq\left|\mathcal{X}\right|+2$
and $\left|\mathcal{U}_{2}\right|\leq\left(\left|\mathcal{X}\right|+1\right)^{2}$;
iii) there exist functions $\varphi_{1}$ and $\varphi_{2}$ such
that $\mathbb{E}\left[d\left(X,\varphi_{1}\left(U_{1}\right)\right)\right]\leq D_{1}$
and $\mathbb{E}\left[d\left(X,\varphi_{2}\left(U_{1},U_{2},Y\right)\right)\right]\leq D_{2}.$ 
\end{thm}

\subsection{Doubly Symmetric Binary Sources }

\begin{figure}
\begin{centering}
\includegraphics[bb=0bp 200bp 595bp 620bp,clip,scale=0.6]{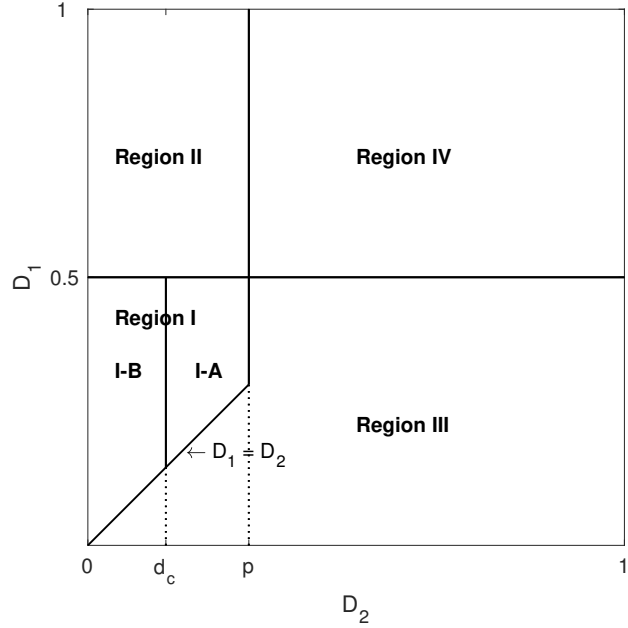}
\par\end{centering}
\centering{}\caption{Illustration of the HBRDF regions for DSBS, where $d_{c}$ is the
critical distortion. \label{fig:RegionPlotDSBS} }
\end{figure}

Let $X$ be a binary DMS, i.e., $\mathcal{X}=\left\{ 0,1\right\} $,
with uniform distribution. The binary side information is specified
by $Y=X\oplus Z$, where $Z$ is an independent Bernoulli random variable
with $P_{Z}\left(z=1\right)=p<0.5$, and $\oplus$ denotes modulo
two addition.

The HBRDF for DSBS can be characterized over four regions \cite{Kerpez1987HB}.
Region I ($0\leq D_{1}<0.5$ and $0\leq D_{2}<\min\left(D_{1},p\right)$)
is a non-degenerate region, and $R_{HB}(D_{1},D_{2})$ is a function
of $D_{1}$ and $D_{2}$; Region II ($D_{1}\geq0.5$ and $0\leq D_{2}\leq p$)
is a degenerate region as the Heegard-Berger problem boils down to
the Wyner-Ziv problem for the second decoder; Region III ($0\leq D_{1}\leq0.5$
and $D_{2}\geq\min\left(D_{1},p\right)$) is also degenerate since
the problem boils down to the standard lossy compression problem for
the first decoder; Region IV ($D_{1}>0.5$ and $D_{2}>p$) can be
trivially achieved without coding. These four regions are depicted
in Fig. \ref{fig:RegionPlotDSBS}. Note that, the HBRDF in the degenerate
Regions II and III can be achieved by using polar codes as described
in \cite{polarchannelandsource}. Here we focus on the non-degenerate
Region I. 

The explicit calculation of HBRDF for DSBS in Region I is given in
\cite{Tian2006HB} as follows: 

Define the function 
\[
\begin{aligned} & S_{D_{1}}\left(\alpha,\mu,\theta,\theta_{1}\right)\triangleq1-h\left(D_{1}*p\right)+\left(\theta-\theta_{1}\right)G\left(\alpha\right)+\theta_{1}G\left(\mu\right)+\left(1-\theta\right)G\left(\gamma\right),\end{aligned}
\]
 where 
\[
\begin{array}{cc}
\gamma\triangleq & \begin{cases}
\frac{D_{1}-\left(\theta-\theta_{1}\right)\left(1-\alpha\right)-\theta_{1}\mathbf{\mu}}{1-\theta} & \theta\neq1\\
0.5 & \theta=1
\end{cases}\end{array},
\]
 on the domain $0\leq\theta_{1}\leq\theta\leq1,\textrm{ }0\leq\alpha,\mu\leq p,\textrm{ }p\leq\gamma\leq1-p.$

The following theorem characterizes the HBRDF in Region I.
\begin{thm}
\label{thm:HBRD_region_I}\cite[Theorem 2]{Tian2006HB} For $0\leq D_{1}<0.5$
and $0\leq D_{2}<\min\left(D_{1},p\right)$, we have $R_{HB}\left(D_{1},D_{2}\right)=\min S_{D_{1}}\left(\alpha,\mu,\theta,\theta_{1}\right)$,
where the minimization is over all $\theta_{1}$, $\theta$, $\alpha$
and $\mu$ variables that satisfy $0\leq\theta_{1}\leq\theta\leq1$,
$0\leq\alpha,\mu\leq p$ and $\left(\theta-\theta_{1}\right)\alpha+\theta_{1}\mu+\left(1-\theta\right)p=D_{2}$. 
\end{thm}
The corresponding forward test channel structure is also given in
\cite{Tian2006HB}, reproduced in Table \ref{tab:Joint-distribution}.
It constructs random variables with joint distribution $P_{X,U_{1},U_{2}}\left(x,u_{1},u_{2}\right)$,
which satisfy $I\left(X;U_{1}\right)+I\left(X;U_{2}|U_{1},Y\right)=S_{D_{1}}\left(\alpha,\mu,\theta,\theta_{1}\right)$. 

\begin{table}[t]
\begin{centering}
\begin{tabular}{|c||c|c|c|c|}
\hline 
 & $\left(u_{1},x\right)=(0,0)$ & $\left(u_{1},x\right)=(0,1)$ & $\left(u_{1},x\right)=(1,0)$ & $\left(u_{1},x\right)=(1,1)$\tabularnewline
\hline 
\hline 
$u_{2}=0$ & $\frac{1}{2}\theta_{1}\left(1-\mu\right)$ & $\frac{1}{2}\theta_{1}\mu$ & $\frac{1}{2}\left(\theta-\theta_{1}\right)\left(1-\alpha\right)$ & $\frac{1}{2}\left(\theta-\theta_{1}\right)\alpha$\tabularnewline
\hline 
$u_{2}=1$ & $\frac{1}{2}\left(\theta-\theta_{1}\right)\alpha$ & $\frac{1}{2}\left(\theta-\theta_{1}\right)\left(1-\alpha\right)$ & $\frac{1}{2}\theta_{1}\mu$ & $\frac{1}{2}\theta_{1}\left(1-\mu\right)$\tabularnewline
\hline 
$u_{2}=2$ & $\frac{1}{2}\left(1-\theta\right)\left(1-\gamma\right)$ & $\frac{1}{2}\left(1-\theta\right)\gamma$ & $\frac{1}{2}\left(1-\theta\right)\gamma$ & $\frac{1}{2}\left(1-\theta\right)\left(1-\gamma\right)$\tabularnewline
\hline 
\end{tabular}
\par\end{centering}
\begin{centering}
\vspace{0.2cm}
\caption{Joint distribution $P_{X,U_{1},U_{2}}\left(x,u_{1},u_{2}\right)$
\cite{Tian2006HB}.\label{tab:Joint-distribution}}
\par\end{centering}
\vspace{-0.8cm}
\end{table}

Next, we recall the function $G\left(u\right)\triangleq h\left(p*u\right)-h(u)$
from \cite{Wyner1976Ziv}, defined over the domain $0\leq u\leq1$,
where $h(u)$ is the binary entropy function $h(u)\triangleq-u\log u-(1-u)\log\left(1-u\right)$,
and $p*u$ is the binary convolution for $0\leq p,u\leq1$, defined
as $p*u\triangleq p\left(1-u\right)+u\left(1-p\right)$. Then, recall
the definition of the critical distortion, $d_{c}$, in the Wyner-Ziv
problem for DSBS \cite{Wyner1976Ziv}, for which $\frac{G\left(d_{c}\right)}{d_{c}-p}=G'\left(d_{c}\right)$.

\begin{figure}[tp]
\begin{centering}
\includegraphics[scale=0.55]{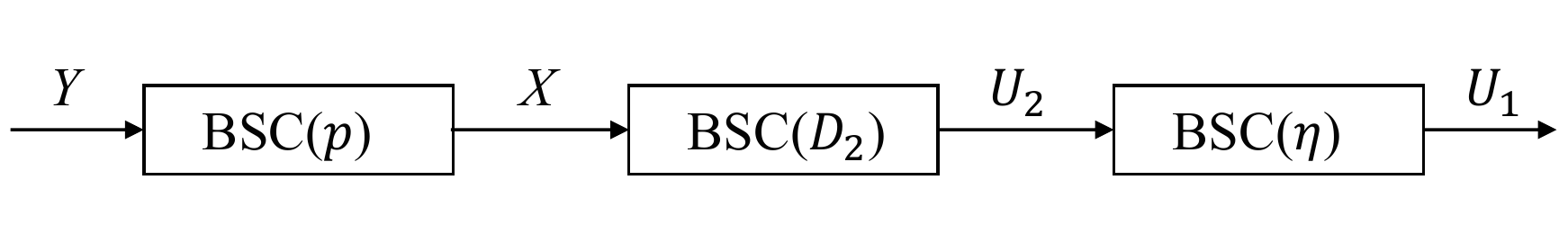}
\par\end{centering}
\vspace{-0.4cm}

\caption{The optimal forward test channel for Region I-B. The crossover probability
$\eta$ for the BSC between $U_{2}$ and $U_{1}$ satisfies $D_{2}*\eta=D_{1}$.
\label{fig:ForwardTestChannel}}

\vspace{-0.4cm}
\end{figure}

The following corollary from \cite{Tian2006HB} specifies an explicit
the characterization of HBRDF for DSBS in Region I-B in Fig. \ref{fig:RegionPlotDSBS}
specified by $D_{2}\leq\min\left(d_{c},D_{1}\right)$ and $D_{1}\leq0.5$.
\begin{cor}
(\cite[Corollary 2]{Tian2006HB}) \label{cor:R_HB_bound}For distortion
pairs $\left(D_{1},D_{2}\right)$ satisfying $D_{1}\leq0.5$ and $D_{2}\leq\min\left(d_{c},D_{1}\right)$
(i.e., Region I-B in Fig. \ref{fig:RegionPlotDSBS}), we have 
\begin{equation}
R_{HB}\left(D_{1},D_{2}\right)=1-h\left(D_{1}*p\right)+G\left(D_{2}\right).\label{eq:Rhb_DSBS}
\end{equation}
\end{cor}
From \cite{Tian2006HB}, the optimal forward test channel for Region
I-B is given as a cascade of two binary symmetric channels (BSCs),
as depicted in Fig. \ref{fig:ForwardTestChannel}. 

In Section \ref{sec:Polar codes for DSBS}, we first propose a polar
code design that achieves the HBRDF in Region I-B for DSBSs. We then
provide a general polar code construction achieving the HBRDF in the
entire Region I.

\subsection{Gaussian Sources \label{subsec:The-Gaussian-Case}}

Suppose $Y=X+Z$, where $X$ and $Z$ are independent (zero-mean)
Gaussian random variables with variances $\sigma_{X}^{2}$ and $\sigma_{Z}^{2}$,
respectively, i.e., $X\sim\mathcal{N}\left(0,\sigma_{X}^{2}\right)$
and $Z\sim\mathcal{N}\left(0,\sigma_{Z}^{2}\right)$. The explicit
expression for $R_{HB}\left(D_{1},D_{2}\right)$ in this case is given
in \cite{Heegard1985absent}. The optimal test channels are given
by $X=U_{1}+Z_{1}$ and $X=U_{2}+Z_{2}$, where $Z$, $Z_{1}$ and
$Z_{2}$ are independent zero-mean Gaussian random variables. We have
$Z_{i}\sim\mathcal{N}\left(0,D_{i}\right),\textrm{ }i=1,2$. 

For $D_{1}\leq\sigma_{X}^{2}$ and $D_{2}\geq\frac{D_{1}\sigma_{Z}^{2}}{D_{1}+\sigma_{Z}^{2}}$,
the problem degenerates into a classical lossy compression problem
for Decoder 1, and the HBRDF is given by $R_{HB}\left(D_{1},D_{2}\right)=\frac{1}{2}\log\left(\frac{\sigma_{X}^{2}}{D_{1}}\right)$.
For $D_{1}>\sigma_{X}^{2}$ and $D_{2}\leq\frac{D_{1}\sigma_{Z}^{2}}{D_{1}+\sigma_{Z}^{2}}$,
the problem degenerates into a Wyner-Ziv coding problem for Decoder
2, and we have $R_{HB}\left(D_{1},D_{2}\right)=\frac{1}{2}\log\left(\frac{\sigma_{X}^{2}\sigma_{Z}^{2}}{D_{2}\left(\sigma_{X}^{2}+\sigma_{Z}^{2}\right)}\right)$.
The region specified by $D_{1}>\sigma_{X}^{2}$ and $D_{2}\geq\frac{D_{1}\sigma_{Z}^{2}}{D_{1}+\sigma_{Z}^{2}}$
requires no coding. Polar lattice codes that meet the classical and
Wyner-Ziv rate-distortion functions for Gaussian sources, introduced
in \cite{PolarlatticeQZ}, can be used to achieve the HBRDF in these
degenerate regions. The only non-degenerate distortion region is specified
by $D_{1}\leq\sigma_{X}^{2}$ and $D_{2}\leq\frac{D_{1}\sigma_{Z}^{2}}{D_{1}+\sigma_{Z}^{2}}$,
and the HBRDF in this region is given by \cite{Heegard1985absent}:
{\allowdisplaybreaks 
\begin{equation}
\begin{aligned} & R_{HB}\left(D_{1},D_{2}\right)=\frac{1}{2}\log\left(\frac{\sigma_{X}^{2}\sigma_{Z}^{2}}{D_{2}\left(D_{1}+\sigma_{Z}^{2}\right)}\right).\end{aligned}
\label{eq:Rhb_boundGaussian}
\end{equation}
}We will focus on the construction of polar lattice codes that achieve
the HBRDF in (\ref{eq:Rhb_boundGaussian}) in Section \ref{sec:Polar-Lattices-forGaussianSources}.

\section{Polar Codes for DSBS \label{sec:Polar codes for DSBS}}

In this section, we present a construction of polar codes that achieves
$R_{HB}\left(D_{1},D_{2}\right)$ for DSBS in Region I. First, we
give a brief overview of polar codes. 

Let $G_{2}\triangleq\ensuremath{\left[\begin{smallmatrix}1 & 0\\
1 & 1
\end{smallmatrix}\right]}$, and define $G_{N}\triangleq G_{2}^{\otimes n}$ as the generator
matrix of polar codes with length $N=2^{n}$, where \textquoteleft $\otimes$\textquoteright{}
denotes the Kronecker product. A polar code $\mathcal{C}_{N}\left(\mathcal{F},u_{\mathcal{F}}\right)$
\cite{polarchannelandsource} is a linear code defined by $\mathcal{C}_{N}\left(\mathcal{F},u_{\mathcal{F}}\right)=\left\{ v^{1:N}G_{N}:v_{\mathcal{F}}=u_{\mathcal{F}},v_{\mathcal{F}^{c}}\in\left\{ 0,1\right\} ^{\left|\mathcal{F}^{c}\right|}\right\} ,$
for any $\mathcal{F}\subseteq\left[N\right]$ and $u_{\mathcal{F}}\in\left\{ 0,1\right\} ^{\left|\mathcal{F}\right|}$,
where $\mathcal{F}$ is referred to as the frozen set. The code $\mathcal{C}_{N}\left(\mathcal{F},u_{\mathcal{F}}\right)$
is constructed by fixing $u_{\mathcal{F}}$ and varying the values
in $\mathcal{F}^{c}$. Moreover, the frozen set can be determined
by the Bhattacharyya parameter \cite{polarchannelandsource}. For
a binary memoryless asymmetric channel with input $X\in\mathcal{X}=\left\{ 0,1\right\} $
and output $Y\in\mathcal{Y}$, the Bhattacharyya parameter $Z$ is
defined as $Z\left(X|Y\right)\triangleq2\sum_{y}\sqrt{P_{X,Y}\left(0,y\right)P_{X,Y}\left(1,y\right)}$. 

\subsection{Polar Code Construction for Region I-B \label{subsec:For region I-B}}

We observe from the proof of Theorem \ref{thm:HBRD_region_I} in \cite{Tian2006HB}
that the auxiliary random variable $U_{1}$ can be considered as the
output of a BSC with crossover probability $D_{1}$ and input $X$.
Therefore, as for Region I-B, the minimum rate for Decoder 1 to achieve
the target distortion $D_{1}$ is $R_{1}=I\left(U_{1};X\right)=1-h\left(D_{1}\right).$
It is shown in \cite[Theorem 3]{KoradaSource} that polar codes can
achieve the rate-distortion function of binary symmetric sources.
An explicit code construction is also provided in \cite{KoradaSource}.
Considering the source sequence $X^{1:N}$ as $N$ independent and
identically distributed (i.i.d) copies of $X$, we know from \cite[Theorem 3]{KoradaSource}
that Decoder 1 can recover a reconstruction $\hat{X}_{1}^{1:N}$ that
is asymptotically close to $U_{1}^{1:N}$ as $N$ becomes sufficiently
large. Therefore, we assume that both Decoder 1 and Decoder 2 can
obtain $U_{1}^{1:N}$ in the following. 

Decoder 2 observes the side information $Y^{1:N}$, in addition to
$U_{1}^{1:N}$ that can be reconstructed using the same method as
Decoder 1. Hence, both $Y^{1:N}$ and $U_{1}^{1:N}$ can be considered
as side information for Decoder 2 to achieve distortion $D_{2}$.
Therefore, the problem at Decoder 2 is very similar to Wyner-Ziv coding
except that the decoder observes extra side information.

Recall that achieving the Wyner-Ziv rate-distortion function using
polar codes is based on the nested code structure proposed in \cite{polarchannelandsource}.
Consider the Wyner-Ziv problem consisting of compressing a source
$X^{1:N}$ in the presence of correlated side information $Y^{1:N}$
using polar codes, where $X$ and $Y$ are DSBS. The code $\mathcal{C}_{s}$
with corresponding frozen set $\mathcal{F}_{s}$ is designed to be
a good source code for distortion $D_{2}$. Further, the code $\mathcal{C}_{c}$
with corresponding frozen set $\mathcal{F}_{c}$ is designed to be
a good channel code for BSC$\left(D_{2}*p\right)$. It has been shown
in \cite{polarchannelandsource} that $\mathcal{F}_{c}\supseteq\mathcal{F}_{s}$,
because the test channel BSC$\left(D_{2}*p\right)$ is degraded with
respect to BSC$\left(D_{2}\right)$. In this case, the encoder transmits
to the decoder the sub-vector that belongs to the index set $\mathcal{F}_{c}$\textbackslash{}$\mathcal{F}_{s}$.
The optimality of this scheme is proven in \cite{polarchannelandsource}.

Similarly, the optimal rate-distortion performance for Decoder 2 in
the Heegard-Berger problem can also be achieved by using nested polar
codes. For $\left(U_{1},U_{2}\right)\in\mathcal{P}\left(D_{1},D_{2}\right)$,
we have
\begin{equation}
\begin{aligned}I\left(X;U_{2}|U_{1},Y\right) & =I\left(U_{2};X,U_{1},Y\right)-I\left(U_{2};Y,U_{1}\right)\\
 & =I\left(U_{2};X,U_{1}\right)-I\left(U_{2};Y,U_{1}\right).
\end{aligned}
\label{eq:rate of decoder 2 binary}
\end{equation}
The second equality holds since $Y\leftrightarrow X\leftrightarrow\left(U_{1},U_{2}\right)$
form a Markov chain. Motivated by $\left(\ref{eq:rate of decoder 2 binary}\right)$,
the code $\mathcal{C}_{s_{2}}$ with corresponding frozen set $\mathcal{F}_{s_{2}}$
is designed to be a good source code for the source pair $\left(X^{1:N},U_{1}^{1:N}\right)$
with reconstruction $U_{2}^{1:N}$. $T_{s}$ denotes the test channel
for this source code. Additionally, the code $\mathcal{C}_{c_{2}}$
with corresponding frozen set $\mathcal{F}_{c_{2}}$ is designed to
be a good channel code for the test channel $T_{c}$ with input $U_{2}^{1:N}$
and output $\left(Y^{1:N},U_{1}^{1:N}\right)$. According to \cite[Lemma 4.7]{polarchannelandsource},
in order to show the nested structure between $\mathcal{C}_{s_{2}}$
and $\mathcal{C}_{c_{2}}$, it is sufficient to show that $T_{c}$
is stochastically degraded with respect to $T_{s}$.
\begin{defn}
(Channel Degradation \cite{polarchannelandsource}). Let $T_{1}:\mathcal{U}\rightarrow\mathcal{Y}$
and $T_{2}:\mathcal{U}\rightarrow\mathcal{X}$ be two binary discrete
memoryless channels. We say that $T_{1}$ is stochastically degraded
with respect to $T_{2}$, if there exists a discrete memoryless channel
$T:\mathcal{X}\rightarrow\mathcal{Y}$ such that 
\[
P_{Y|U}(y|u)=\sum_{x\in\mathcal{X}}P_{X|U}(x|u)P_{Y|X}(y|x).
\]
\end{defn}
\begin{prop}
\label{prop: test channel degradation}$T_{c}:U_{2}\rightarrow\left(Y,U_{1}\right)$
is stochastically degraded with respect to $T_{s}:U_{2}\rightarrow\left(X,U_{1}\right)$,
if the random variables $\left(X,Y,U_{1},U_{2}\right)$ agree with
the forward test channel as shown in Fig. \ref{fig:ForwardTestChannel}.
\end{prop}
\begin{IEEEproof}
From the test channel structure in Fig. \ref{fig:ForwardTestChannel},
$Y\leftrightarrow X\leftrightarrow U_{2}\leftrightarrow U_{1}$ form
a Markov chain. By definition, we have $P_{X,U_{1}|U_{2}}\left(x,u_{1}|u_{2}\right)=P_{X|U_{2}}\left(x|u_{2}\right)P_{U_{1}|U_{2}}\left(u_{1}|u_{2}\right).$
We also have 
\[
\begin{aligned}P_{Y,U_{1}|U_{2}}\left(y,u_{1}|u_{2}\right) & =P_{Y|U_{2}}\left(y|u_{2}\right)P_{U_{1}|U_{2}}\left(u_{1}|u_{2}\right)\\
 & =\sum_{x}P_{X,Y|U_{2}}\left(x,y|u_{2}\right)P_{U_{1}|U_{2}}\left(u_{1}|u_{2}\right)\\
 & =\sum_{x}P_{X|U_{2}}\left(x|u_{2}\right)P_{Y|X,U_{2}}\left(y|x,u_{2}\right)P_{U_{1}|U_{2}}\left(u_{1}|u_{2}\right)\\
 & =\sum_{x}P_{X,U_{1}|U_{2}}\left(x,u_{1}|u_{2}\right)P_{Y|X}\left(y|x\right),
\end{aligned}
\]
completing the proof.
\end{IEEEproof}
Therefore, we can claim that $\mathcal{F}_{c_{2}}\supseteq\mathcal{F}_{s_{2}}$
by \cite[Lemma 4.7]{polarchannelandsource}, and rather than sending
the entire vector that belongs to the index set $\mathcal{F}_{s_{2}}^{c}$,
the encoder sends only the sub-vector that belongs to $\mathcal{F}_{c_{2}}$\textbackslash{}$\mathcal{F}_{s_{2}}$
to Decoder 2, since Decoder 2 can extract some information on $U_{2}^{1:N}$
from the available side information $\left(U_{1}^{1:N},Y^{1:N}\right)$.
As a result, the polar code construction for the Heegard-Berger problem
in Region I-B is given as follows:

\textit{\uline{Encoding:}} The encoder first applies lossy compression
to source sequence $X^{1:N}$ with reconstruction $U_{1}^{1:N}$ and
corresponding average distortion $D_{1}$. We construct the code $\mathcal{C}_{s_{1}}=\mathcal{C}_{N}\left(\mathcal{F}_{s_{1}},\bar{0}\right)=\left\{ w^{1:N}G_{N}:w_{\mathcal{F}_{s_{1}}}=\bar{0},w_{\mathcal{F}_{s_{1}}^{c}}\in\left\{ 0,1\right\} ^{\left|\mathcal{F}_{s_{1}}^{c}\right|}\right\} $,
and the encoder transmits the compressed sequence $w_{\mathcal{F}_{s_{1}}^{c}}$
to the decoders. The encoder is also able to recover $U_{1}^{1:N}$
from $\mathcal{C}_{s_{1}}$. Next, the encoder applies lossy compression
jointly for sources $\left(X^{1:N},U_{1}^{1:N}\right)$ with reconstruction
$U_{2}^{1:N}$ and target distortion $D_{2}$ and $d\left(U_{1},U_{2}\right)=\eta$.
We then construct $\mathcal{C}_{s_{2}}=\mathcal{C}_{N}\left(\mathcal{F}_{s_{2}},\bar{0}\right)$.
Finally, the encoder applies channel coding to the symmetric test
channel $T_{c}$ with input $U_{2}^{1:N}$ and output $\left(Y^{1:N},U_{1}^{1:N}\right)$.
We derive $\mathcal{C}_{c_{2}}=\mathcal{C}_{N}\left(\mathcal{F}_{c_{2}},u_{\mathcal{F}_{c_{2}}}\left(\bar{v}\right)\right)$,
where $u_{\mathcal{F}_{c_{2}}}\left(\bar{v}\right)$ is defined by
$u_{\mathcal{F}_{s_{2}}}=\bar{0}$ and $u_{\mathcal{F}_{c_{2}}\backslash\mathcal{F}_{s_{2}}}=\bar{v}$
for $\bar{v}\in\{0,1\}^{\left|\mathcal{F}_{c_{2}}\backslash\mathcal{F}_{s_{2}}\right|}$.
The encoder sends the sub-vector $u_{\mathcal{F}_{c_{2}}\backslash\mathcal{F}_{s_{2}}}$
to the decoders.

\textit{\uline{Decoding:}} Decoder 1 receives $w_{\mathcal{F}_{s_{1}}^{c}}$
and outputs the reconstruction sequence $u_{1}^{1:N}=w^{1:N}G_{N}$.
Decoder 2 receives $u_{\mathcal{F}_{c_{2}}\backslash\mathcal{F}_{s_{2}}}$,
and hence, it can derive $u_{\mathcal{F}_{c_{2}}}$. Moreover, Decoder
2 can also recover $U_{1}^{1:N}$ from $w_{\mathcal{F}_{s_{1}}^{c}}$.
Decoder 2 applies the successive cancellation (SC) decoding algorithm
to obtain the codeword $U_{2}^{1:N}$ from the realizations of $\left(Y^{1:N},U_{1}^{1:N}\right)$.

Next we present the rates that can be achieved by the proposed scheme.
From the polarization theorem for lossy source coding in \cite{KoradaSource},
we know that reliable decoding at Decoder 1 will be achieved with
high probability if $\frac{\left|\mathcal{F}_{s_{1}}^{c}\right|}{N}\xrightarrow{N\rightarrow\infty}I\left(U_{1};X\right)=1-h\left(D_{1}\right)$. 

From the polarization theorems for source and channel coding \cite{polarchannelandsource},
the code rate required for reliable decoding at Decoder 2 can be derived
by 
\[
\begin{aligned}\frac{\left|\mathcal{F}_{c_{2}}\right|-\left|\mathcal{F}_{s_{2}}\right|}{N}\xrightarrow{N\rightarrow\infty} & I\left(U_{2};X,U_{1}\right)-I\left(U_{2};Y,U_{1}\right)\\
 & =I\left(U_{2};X\right)+I\left(U_{1};X,U_{2}\right)-I\left(U_{1};X\right)-I\left(U_{2};Y\right)-I\left(U_{1};Y,U_{2}\right)+I\left(U_{1};Y\right)\\
 & =G\left(D_{2}\right)-G\left(D_{1}\right).\text{}
\end{aligned}
\]
 Therefore, the total rate will be asymptotically given by $\frac{\left|\mathcal{F}_{s_{1}}^{c}\right|+\left|\mathcal{F}_{c_{2}}\right|-\left|\mathcal{F}_{s_{2}}\right|}{N}\xrightarrow{N\rightarrow\infty}1-h\left(D_{1}*p\right)+G\left(D_{2}\right),$
for Region I-B.

\begin{figure}
\vspace{-0.4cm}

\begin{centering}
\includegraphics[scale=0.65]{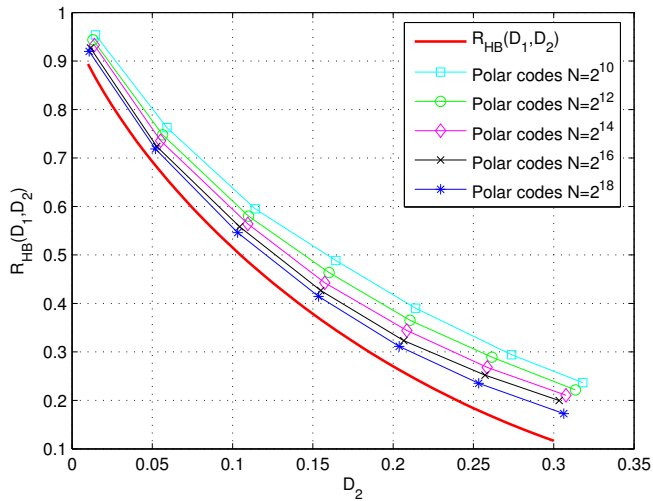}
\par\end{centering}
\vspace{-0.4cm}

\caption{Simulation performance of $R_{HB}\left(D_{1},D_{2}\right)$ corresponding
to $D_{2}$ for Region I-B. \label{fig:Simulation-results-DSBS} }
\end{figure}

Furthermore, according to \cite{polarchannelandsource,KoradaSource},
the expected distortions asymptotically approach the target values
$D_{1}$ and $D_{2}$ at Decoders 1 and 2, respectively, as $N$ becomes
sufficiently large. The encoding and decoding complexity of this scheme
is $O\left(N\log N\right)$.

Note that, in our scheme, the performance of Decoder 2 is more challenging
than that of Decoder 1. Thus, the simulation is conducted by fixing
$D_{1}=0.35$, $p=0.4$, and varying $D_{2}\in\left(0,\min\left(d_{c},D_{1}\right)\right)$.
These settings satisfy the requirements for Region I-B. The performance
curves are shown in Fig. \ref{fig:Simulation-results-DSBS} for $n=10,12,14,16,18$.
It shows that the performances achieved by polar codes approaches
the HBRDF as $n$ increases.

\subsection{Coding Scheme for Entire Region I \label{subsec:Proposed-Coding-Scheme-Region I}}

As Theorem \ref{thm:HBRD_region_I} defines the $R_{HB}\left(D_{1},D_{2}\right)$
for the entire Region I, we now present a coding scheme that can achieve
the HBRDF for the entire Region I. Note that $R_{HB}\left(D_{1},D_{2}\right)$
of Region I-B can be explicitly calculated by Corollary \ref{cor:R_HB_bound}.
Therefore, we can also achieve Region I-B straightforwardly as shown
in Section \ref{subsec:For region I-B}.

From the optimal test channel structure shown in Table \ref{tab:Joint-distribution},
$U_{2}$ is a ternary random variable, i.e., $\mathcal{U}_{2}=\left\{ 0,1,2\right\} $.
Therefore, we express $U_{2}$ as two binary random variables $U_{a}$
and $U_{b}$, where $U_{2}=2U_{b}+U_{a}$, i.e., $\left(U_{a},U_{b}\right)\in\left\{ 00,10,01\right\} $.
For Decoder 1, we can apply the same scheme specified in the previous
subsection to achieve $D_{1}$. Again, $U_{1}^{1:N}$ and $Y^{1:N}$
can be considered as side information for Decoder 2. Then, the rate
required to transmit $U_{2}^{1:N}$ is evaluated as $I\left(X;U_{2}|U_{1},Y\right)=I\left(X;U_{a},U_{b}|U_{1},Y\right)=I\left(X;U_{a}|U_{1},Y\right)+I\left(X;U_{b}|U_{1},U_{a},Y\right)$.
Accordingly we can design two separate coding schemes to achieve the
rates $I\left(X;U_{a}|U_{1},Y\right)$ and $I\left(X;U_{b}|U_{1},U_{a},Y\right)$,
respectively. 

Since $Y\leftrightarrow X\leftrightarrow(U_{1},U_{a},U_{b})$ form
a Markov chain, we have $I\left(X;U_{a}|U_{1},Y\right)=I\left(U_{a};X,U_{1}\right)-I\left(U_{a};Y,U_{1}\right)$,
and the test channel $T_{C_{a}}:U_{a}\rightarrow\left(Y,U_{1}\right)$
is degraded with respect to $T_{S_{a}}:U_{a}\rightarrow\left(X,U_{1}\right)$.
We can observe from Table \ref{tab:Joint-distribution} that $U_{a}$
and $U_{b}$ can be nonuniform. 

Let $K^{1:N}=U_{a}^{1:N}G_{N}$, and for $0<\beta<0.5$, the frozen
set $\mathcal{F}_{S_{a}}\left(\mathcal{F}_{C_{a}}\right)$, the information
set $\mathcal{I}_{S_{a}}\left(\mathcal{I}_{C_{a}}\right)$, and the
shaping set $\mathcal{S}_{S_{a}}\left(\mathcal{S}_{C_{a}}\right)$
can be identified as 

\begin{equation}
\hspace{-2bp}\begin{split} & \mathcal{F}_{S_{a}}=\left\{ i\in[N]\hspace{-3bp}:\hspace{-3bp}Z\left(K^{i}|K^{1:i-1},X^{1:N},U_{1}^{1:N}\right)\geq1-2^{-N^{\beta}}\right\} \\
 & \mathcal{I}_{S_{a}}=\left\{ i\in[N]\hspace{-3bp}:\hspace{-3bp}Z\left(K^{i}|K^{1:i-1},X^{1:N},U_{1}^{1:N}\right)<1-2^{-N^{\beta}}\right\} \cap\left\{ i\in[N]\hspace{-3bp}:\hspace{-3bp}Z\left(K^{i}|K^{1:i-1}\right)>2^{-N^{\beta}}\right\} \\
 & \mathcal{S}_{S_{a}}=\left\{ i\in[N]\hspace{-3bp}:\hspace{-3bp}Z\left(K^{i}|K^{1:i-1}\right)\leq2^{-N^{\beta}}\right\} ,\\
 & \mathcal{F}_{C_{a}}=\left\{ i\in[N]\hspace{-3bp}:\hspace{-3bp}Z\left(K^{i}|K^{1:i-1},Y^{1:N},U_{1}^{1:N}\right)\geq1-2^{-N^{\beta}}\right\} \\
 & \mathcal{I}_{C_{a}}=\left\{ i\in[N]\hspace{-3bp}:\hspace{-3bp}Z\left(K^{i}|K^{1:i-1},Y^{1:N},U_{1}^{1:N}\right)\leq2^{-N^{\beta}}\right\} \cap\left\{ i\in[N]\hspace{-3bp}:\hspace{-3bp}Z\left(K^{i}|K^{1:i-1}\right)\geq1-2^{-N^{\beta}}\right\} \\
 & \mathcal{S}_{C_{a}}=\left\{ i\in[N]\hspace{-3bp}:\hspace{-3bp}Z\left(K^{i}|K^{1:i-1}\right)<1-2^{-N^{\beta}}\right\} \\
 & \hspace{40bp}\cup\left\{ i\in[N]\hspace{-3bp}:\hspace{-5bp}\textrm{ }2^{-N^{\beta}\hspace{-1bp}}\hspace{-0.5bp}<Z\left(K^{i}|K^{1:i-1},Y^{1:N},U_{1}^{1:N}\right)<\hspace{-1bp}1-2^{-N^{\beta}}\hspace{-2bp}\right\} .
\end{split}
\label{eq:SetDivision}
\end{equation}

By \cite[Lemma 4.7]{polarchannelandsource} and channel degradation,
we have $\mathcal{F}_{S_{a}}\subseteq\mathcal{F}_{C_{a}}$, $\mathcal{I}_{C_{a}}\subseteq\mathcal{I}_{S_{a}}$
and $\mathcal{S}_{S_{a}}\subseteq\mathcal{S}_{C_{a}}$. In addition,
we observe that $\mathcal{S}_{C_{a}}\backslash\mathcal{S}_{S_{a}}$
can be written as 
\[
\begin{aligned}\mathcal{S}_{C_{a}}\backslash\mathcal{S}_{S_{a}}= & \left\{ i\in[N]\hspace{-3bp}:\hspace{-3bp}2^{-N^{\beta}}<Z\left(K^{i}|K^{1:i-1}\right)<1-2^{-N^{\beta}}\right\} \cup\\
 & \left\{ i\in[N]\hspace{-3bp}:\hspace{-5bp}\textrm{ }2^{-N^{\beta}\hspace{-1bp}}\hspace{-0.5bp}<Z\left(K^{i}|K^{1:i-1},Y^{1:N},U_{1}^{1:N}\right)<\hspace{-1bp}1-2^{-N^{\beta}}\hspace{-2bp}\right\} ,
\end{aligned}
\]
therefore, the proportion $\frac{\left|\mathcal{S}_{C_{a}}\backslash\mathcal{S}_{S_{a}}\right|}{N}\rightarrow0$,
as $N\rightarrow\infty$. 

\textit{\uline{Encoding:}} The encoder first applies lossy compression
to $X^{1:N}$ with target distortion $D_{1}$ to obtain $U_{1}^{1:N}$,
and treats $\left(X^{1:N},U_{1}^{1:N}\right)$ as a joint source sequence
to evaluate $K_{\mathcal{I}_{S_{a}}}$ by randomized rounding with
respect to $P_{K^{i}|K^{1:i-1},X^{1:N},U_{1}^{1:N}}$, i.e., 
\begin{equation}
k^{i}=\begin{cases}
0 & \textrm{w.p. }P_{K^{i}|K^{1:i-1},X^{1:N},U_{1}^{1:N}}\left(0|k^{1:i-1},x^{1:N},u_{1}^{1:N}\right)\\
1 & \textrm{w.p. }P_{K^{i}|K^{1:i-1},X^{1:N},U_{1}^{1:N}}\left(1|k^{1:i-1},x^{1:N},u_{1}^{1:N}\right)
\end{cases}\textrm{if }i\in\mathcal{I}_{S_{a}}\label{eq:randomMapping}
\end{equation}
and 
\begin{equation}
k^{i}=\begin{cases}
\tilde{k}^{i} & \textrm{if }i\in\mathcal{F}_{S_{a}}\\
\arg\max_{k}P_{K^{i}|K^{1:i-1}}\left(k|k^{1:i-1}\right) & \textrm{if }i\in\mathcal{S}_{S_{a}},
\end{cases}\label{eq:MAPdecoding}
\end{equation}
where `w.p.' is an abbreviation of `with probability' in (\ref{eq:randomMapping}),
and $\tilde{k}^{i}$ is chosen uniformly from $\left\{ 0,1\right\} $
and shared between the encoder and the decoders before lossy compression.
Also note that the second formula in (\ref{eq:MAPdecoding}) is in
fact the maximum a posteriori (MAP) decision for $i\in\mathcal{S}_{S_{a}}$.
The encoder sends $K_{\mathcal{I}_{S_{a}}\backslash\mathcal{I}_{C_{a}}}$
to the decoders.

\textit{\uline{Decoding:}} Using the pre-shared $K_{\mathcal{F}_{S_{a}}}$
and received $K_{\mathcal{I}_{S_{a}}\backslash\mathcal{I}_{C_{a}}}$,
Decoder 2 recovers $K_{\mathcal{I}_{C_{a}}}$ and $K_{\mathcal{S}_{S_{a}}}$
from the side information sequences $Y^{1:N}$ and $U_{1}^{1:N}$
by SC decoding algorithm and the MAP rule, respectively. Hence we
obtain $K^{1:N}$. $K_{\mathcal{I}_{C_{a}}\cup\mathcal{S}_{S_{a}}}$
and $K_{\mathcal{S}_{C_{a}}\backslash\mathcal{S}_{S_{a}}}$ can be
recovered with vanishing error probability, since their Bhattacharyya
parameters are arbitrarily small when $N\rightarrow\infty$. Therefore,
the reconstruction is given by $U_{a}^{1:N}=K^{1:N}G_{N}$.
\begin{thm}
\cite[Theorem 2]{PolarlatticeQZ} Let $Q_{K^{1:N},X^{1:N},U_{1}^{1:N}}$
denote the resulting joint distribution derived from (\ref{eq:randomMapping})
and (\ref{eq:MAPdecoding}). Let $P_{K^{1:N},X^{1:N},U_{1}^{1:N}}$
denote the joint distribution as a result of another encoder that
uses (\ref{eq:randomMapping}) for $i\in\left[N\right]$. For any
$\beta'<\beta<0.5$ satisfying (\ref{eq:SetDivision}) and $R_{a}=\frac{\left|\mathcal{I}_{S_{a}}\backslash\mathcal{I}_{C_{a}}\right|}{N}>I\left(X;U_{a}|U_{1},Y\right)$,
we have 
\begin{equation}
\mathbb{V}\left(P_{K^{1:N},X^{1:N},U_{1}^{1:N}},Q_{K^{1:N},X^{1:N},U_{1}^{1:N}}\right)=O\left(2^{-N^{\beta'}}\right),\label{eq:L1_distance}
\end{equation}
where $\mathbb{V}(P_{X},Q_{X})\triangleq\frac{1}{2}\sum_{x}\left|P_{X}(x)-Q_{X}(x)\right|$
denotes the variational distance between distributions $P_{X}$ and
$Q_{X}$.
\end{thm}
Note that, from \cite{aspolarcodes}, $K_{\mathcal{S}_{C_{a}}}$ should
be covered by a pre-shared random mapping to achieve (\ref{eq:L1_distance}).
However, it is shown in \cite[Theorem 2]{PolarlatticeQZ} that replacing
the random mapping with MAP decision for $K_{\mathcal{S}_{S_{a}}}$
preserves the optimality. Thus, we utilize MAP decoder if $i\in\mathcal{S}_{S_{a}}$
in our scheme.

In terms of the second level, the encoder and Decoder 2 first recover
$U_{a}^{1:N}$. Consequently, the encoder treats $\left(X^{1:N},U_{1}^{1:N},U_{a}^{1:N}\right)$
as a joint source, and Decoder 2 treats $\left(Y^{1:N},U_{1}^{1:N},U_{a}^{1:N}\right)$
as a joint side information. Likewise, according to $Y\leftrightarrow X\leftrightarrow(U_{1},U_{a},U_{b})$,
we have $I\left(X;U_{b}|U_{1},U_{a},Y\right)=I\left(U_{b};X,U_{1},U_{a}\right)-I\left(U_{b};Y,U_{1},U_{a}\right)$
and the test channel $T_{C_{b}}:U_{b}\rightarrow\left(Y,U_{1},U_{a}\right)$
is degraded with respect to $T_{S_{b}}:U_{b}\rightarrow\left(X,U_{1},U_{a}\right)$. 

Similar to the first level, let $W^{1:N}=U_{b}^{1:N}G_{N}$ and the
frozen set $\mathcal{F}_{S_{b}}\left(\mathcal{F}_{C_{b}}\right)$,
the information set $\mathcal{I}_{S_{b}}\left(\mathcal{I}_{C_{b}}\right)$,
and the shaping set $\mathcal{S}_{S_{b}}\left(\mathcal{S}_{C_{b}}\right)$
can be adopted from (\ref{eq:SetDivision}) by replacing $K$, $\left(X,U_{1}\right)$
and $\left(Y,U_{1}\right)$ with $W$, $\left(X,U_{1},U_{a}\right)$
and $\left(Y,U_{1},U_{a}\right)$, respectively. As a result, we have
$\mathcal{F}_{S_{b}}\subseteq\mathcal{F}_{C_{b}}$, $\mathcal{I}_{C_{b}}\subseteq\mathcal{I}_{S_{b}}$
and $\mathcal{S}_{S_{b}}\subseteq\mathcal{S}_{C_{b}}$ by \cite[Lemma 4.7]{polarchannelandsource}
and channel degradation. The encoder evaluates $W_{\mathcal{I}_{S_{b}}}$
by randomized rounding with respect to $P_{W^{i}|W^{1:i-1},X^{1:N},U_{1}^{1:N},U_{a}^{1:N}}$,
$W_{\mathcal{F}_{S_{b}}}$ are pre-shared random bits uniformly chosen
from $\left\{ 0,1\right\} $, and $W_{\mathcal{S}_{S_{b}}}$ is determined
by MAP decoder defined as $\arg\max_{w}P_{W^{i}|W^{1:i-1}}\left(w|w^{1:i-1}\right)$.
The encoder sends $W_{\mathcal{I}_{S_{b}}\backslash\mathcal{I}_{C_{b}}}$
to the decoders. Decoder 2 recovers $W_{\mathcal{I}_{C_{b}}\cup\mathcal{S}_{S_{b}}}$
using the pre-shared $W_{\mathcal{F}_{S_{b}}}$ and the side information
$\left(Y^{1:N},U_{1}^{1:N},U_{a}^{1:N}\right)$. Finally, the reconstruction
is given by $U_{b}^{1:N}=W^{1:N}G_{N}$. 

Let $Q_{W^{1:N},X^{1:N},U_{1}^{1:N},U_{a}^{1:N}}$ denote the joint
distribution when the encoder performs compression, according to the
coding scheme presented in the above paragraph. Let $P_{W^{1:N},X^{1:N},U_{1}^{1:N},U_{a}^{1:N}}$
denote the resulting joint distribution of the encoder using randomized
rounding with respect to $P_{W^{i}|W^{1:i-1},X^{1:N},U_{1}^{1:N},U_{a}^{1:N}}$
for all $i\in\left[N\right]$, which means that the encoder dose not
perform compression. Similarly to Theorem 7, for any $\beta'<\beta<0.5$
and $R_{b}=\frac{\left|\mathcal{I}_{S_{b}}\backslash\mathcal{I}_{C_{b}}\right|}{N}>I\left(X;U_{b}|U_{1},U_{a},Y\right)$,
we have 
\begin{equation}
\mathbb{V}\left(P_{W^{1:N},X^{1:N},U_{1}^{1:N},U_{a}^{1:N}},Q_{W^{1:N},X^{1:N},U_{1}^{1:N},U_{a}^{1:N}}\right)=O\left(2^{-N^{\beta'}}\right).\label{eq:L2_distance}
\end{equation}

Note that (\ref{eq:L2_distance}) is based on (\ref{eq:L1_distance}),
and $R_{a}>I\left(X;U_{a}|U_{1},Y\right)$ should be satisfied. Thus,
we have $R_{a}+R_{b}>I\left(X;U_{2}|U_{1},Y\right)$. With regard
to Decoder 2, we can state the following theorem.
\begin{thm}
\label{thm:OptimalityDecoder2} Consider a target distortion $0\leq D_{2}<\min\left(D_{1},p\right)$
for DSBS $X$ when side information $Y$ is available only at the
Decoder 2. For any $0<\beta'<\beta<0.5$, there exists a two-level
polar code with a rate arbitrarily close to $I\left(X;U_{2}|U_{1},Y\right)$,
such that the expected distortion $D_{Q}$ of Decoder 2 satisfies
$D_{Q}\leq D_{2}+O\left(2^{-N^{\beta'}}\right)$.
\end{thm}
\begin{IEEEproof}
See Appendix A. 
\end{IEEEproof}
As for Decoder 1, we know that $U_{1}$ can always be taken as the
output of a BSC with crossover probability $D_{1}$ and input $X$.
Hence, according to \cite[Theorem 3]{KoradaSource} and Theorem \ref{thm:OptimalityDecoder2},
this coding scheme can achieve the optimal HBRDF, as long as the optimal
parameters $\alpha$, $\mu$, $\theta$, and $\theta_{1}$ that achieve
the minimum value of $S_{D_{1}}\left(\alpha,\mu,\theta,\theta_{1}\right)$
can be specified. Finally, we state the achievability of the HBRDF
for DSBS for the entire Region I in the following theorem.
\begin{thm}
Consider target distortions $0\leq D_{1}<0.5$ and $0\leq D_{2}<\min\left(D_{1},p\right)$
for DSBS $X$ when side information $Y$ is available only at Decoder
2. For any $0<\beta'<\beta<0.5$ and any rate $R>\min S_{D_{1}}\left(\alpha,\mu,\theta,\theta_{1}\right)$,
there exist a polar code $\mathcal{C}_{1}$ with rate $R_{1}<I(X,U_{1})$
and a two-level polar code $\mathcal{C}_{2}$ with rate $R_{2}<I\left(X;U_{2}|U_{1},Y\right)$,
with $R_{1}+R_{2}<R$, which together achieve the expected distortions
$D_{1}+O\left(2^{-N^{\beta}}\right)$ at Decoder 1 and $D_{2}+O\left(2^{-N^{\beta'}}\right)$
at Decoder 2, respectively, if $P_{X,U_{1},U_{2}}$ is as given in
Table \ref{tab:Joint-distribution}.
\end{thm}
\begin{IEEEproof}
It has been shown in \cite[Theorem 3]{KoradaSource} that there exists
a polar code with a rate arbitrarily close to $I(X,U_{1})$ that achieves
an expected distortion $D_{1}+O\left(2^{-N^{\beta}}\right)$. Theorem
\ref{thm:OptimalityDecoder2} shows that a two-level polar code with
a rate arbitrarily close to $I\left(X;U_{2}|U_{1},Y\right)$ achieves
$D_{2}+O\left(2^{-N^{\beta'}}\right)$ at Decoder 2. Finally, the
total rate $R_{1}+R_{2}<I(X,U_{1})+I\left(X;U_{2}|U_{1},Y\right)=\min S_{D_{1}}\left(\alpha,\mu,\theta,\theta_{1}\right)$.
The last equality holds if the joint distribution $P_{X,U_{1},U_{2}}$
is the same as that given in Table \ref{tab:Joint-distribution} \cite{Tian2006HB}.
\end{IEEEproof}
Finally, we observe that the encoding and decoding complexity of this
coding scheme is $O\left(N\log N\right)$. 

\section{Polar Lattices for Gaussian Sources \label{sec:Polar-Lattices-forGaussianSources}}

It is shown in \cite{PolarlatticeQZ} that polar lattices achieve
the optimal rate-distortion performance for both the standard and
the Wyner-Ziv compression of Gaussian sources under squared-error
distortion. The Wyner-Ziv problem for the Gaussian case can be solved
by a nested code structure that combines AWGN capacity achieving polar
lattices \cite{polarlatticeJ} and the rate-distortion optimal ones
\cite{PolarlatticeQZ}. Here we show that the HBRDF for the non-degenerate
region specified in (\ref{eq:Rhb_boundGaussian}) can also be achieved
by a similar nested code structure. 

We start with a basic introduction to polar lattices. An $n$-dimensional
lattice is a discrete subgroup of $\mathbb{R}^{n}$ which can be described
by 
\[
\Lambda=\{\lambda=\mathbf{B}z:\textrm{ }z\in\mathbb{Z}^{n}\},
\]
where $\mathbf{B}$ is the full rank generator matrix. For $\sigma>0$
and $c\in\mathbb{R}^{n}$, the Gaussian distribution of variance $\sigma^{2}$
centered at $c$ is defined as 
\[
f_{\sigma,c}(x)=\frac{1}{(\sqrt{2\pi}\sigma)^{n}}e^{-\frac{\|x-c\|^{2}}{2\sigma^{2}}},\:\:x\in\mathbb{R}^{n}.
\]
Let $f_{\sigma,0}(x)=f_{\sigma}(x)$ for short. The $\Lambda$-periodic
function is defined as 
\[
f_{\sigma,\Lambda}(x)=\sum\limits _{\lambda\in\Lambda}f_{\sigma,\lambda}(x)=\frac{1}{(\sqrt{2\pi}\sigma)^{n}}\sum\limits _{\lambda\in\Lambda}e^{-\frac{\|x-\lambda\|^{2}}{2\sigma^{2}}}.
\]
Note that, when $x$ is restricted to the fundamental region $\mathcal{R}\left(\Lambda\right)$,
$f_{\sigma,\Lambda}(x)$ is actually a probability density function
(PDF) of the $\Lambda$-aliased Gaussian noise \cite{forney6}.

The flatness factor of a lattice $\Lambda$ is defined as  
\[
\epsilon_{\Lambda}(\sigma)\triangleq\max\limits _{x\in\mathcal{R}(\Lambda)}|V(\Lambda)f_{\sigma,\Lambda}(x)-1|,
\]
where $V(\Lambda)=\left|\det\left(\mathbf{B}\right)\right|$ denotes
the volume of a fundamental region of $\Lambda$ \cite{forney6}.
It can be interpreted as the maximum variation of $f_{\sigma,\Lambda}(x)$
with respect to the uniform distribution over a fundamental region
of $\Lambda$.

We define the discrete Gaussian distribution over $\Lambda$ centered
at $c$ as the discrete distribution taking values in $\lambda\in\Lambda$
as 
\[
D_{\Lambda,\sigma,c}(\lambda)=\frac{f_{\sigma,c}(\lambda)}{f_{\sigma,c}(\Lambda)},\;\forall\lambda\in\Lambda,
\]
where $f_{\sigma,\mathrm{c}}(\Lambda)=\sum_{\lambda\in\Lambda}f_{\sigma,\mathrm{c}}(\lambda)$.
For convenience, we write $D_{\Lambda,\sigma}=D_{\Lambda,\sigma,0}$.
It has been shown in \cite{LingBel13} that lattice Gaussian distribution
preserves many properties of the continuous Gaussian distribution
when the flatness factor is negligible. To keep the notations simple,
we always set $c=0$ and $n=1$.

A sublattice $\Lambda'\subset\Lambda$ induces a partition (denoted
by $\Lambda/\Lambda'$) of $\Lambda$ into equivalence groups modulo
$\Lambda'$. The order of the partition equals the number of the cosets.
If the order is two, we call this a binary partition. Let $\Lambda\left(\Lambda_{0}\right)/\Lambda_{1}/\cdots/\Lambda_{r-1}/\Lambda'\left(\Lambda_{r}\right)$
for $r\geq1$ be an n-dimensional lattice partition chain. For each
partition $\Lambda_{l-1}/\Lambda_{l}$ $\left(1\leq l\leq r\right)$
a code $\mathcal{C}_{l}$ over $\Lambda_{l-1}/\Lambda_{l}$ selects
a sequence of coset representatives $a_{l}$ in a set $A_{l}$ of
representatives for the cosets of $\Lambda_{l}$. This construction
requires a set of nested linear binary codes $\mathcal{C}_{l}$ with
block length $N$ and dimension of information bits $k_{l}$, and
$\mathcal{C}_{1}\subseteq\mathcal{C}_{2}\cdots\subseteq\mathcal{C}_{r}$.

For the Gaussian Heegard-Berger problem, let $\left(X,Y,Z,Z_{1},Z_{2},U_{1},U_{2}\right)$
be chosen as specified in Section \ref{subsec:The-Gaussian-Case}.
Given $Y$ as the side information for Decoder 2, the HBRDF is given
by (\ref{eq:Rhb_boundGaussian}). To achieve the HBRDF at Decoder
1, we can design a quantization polar lattice for source $X$ with
variance $\sigma_{X}^{2}$ and target distortion $D_{1}$ as in \cite{PolarlatticeQZ}.
As a result, for a target distortion $D_{1}$ and any rate $R_{1}>\frac{1}{2}\log\left(\sigma_{X}^{2}/D_{1}\right)$,
there exists a multilevel polar lattice with rate $R_{1}$, such that
the average distortion is asymptotically close to $D_{1}$ when the
length $N\rightarrow\infty$ and the number of levels $r=O\left(\log\log N\right)$
\cite[Theorem 4]{PolarlatticeQZ}. Therefore, both decoders can recover
$U_{1}^{1:N}$ and $\left(U_{1}^{1:N},Y^{1:N}\right)$ can be regarded
as the side information at Decoder 2. 

\begin{figure}
\begin{centering}
\includegraphics[scale=0.7]{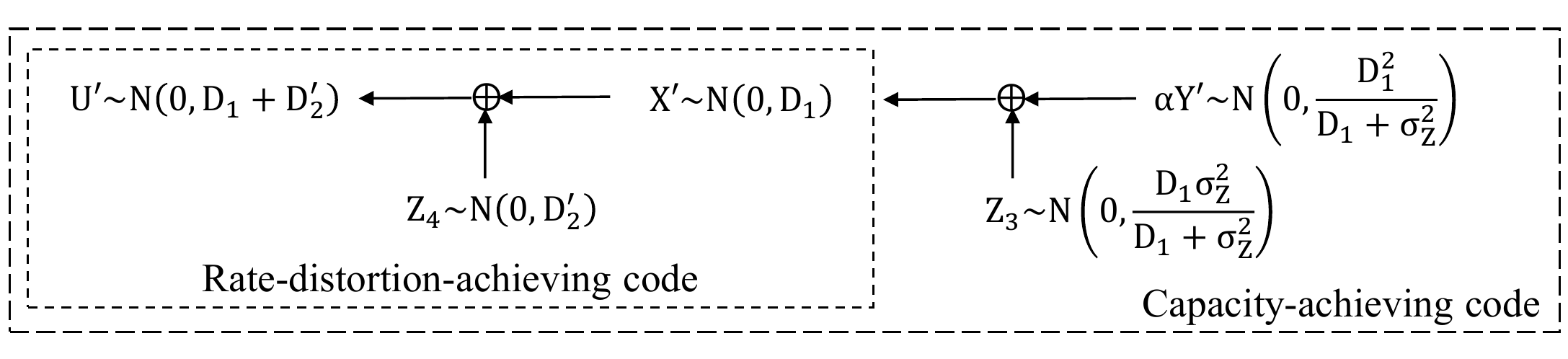}
\par\end{centering}
\caption{A test channel for the Gaussian Heegard-Berger problem for Decoder
2 using a continuous Gaussian $U'$. \label{fig:GaussianTestChannel_1} }

\end{figure}

As for Decoder 2, we first need a code that achieves the rate-distortion
requirement for source $X'\triangleq X-U_{1}$ with Gaussian reconstruction
alphabet $U'$. In fact, $X'=Z_{1}\sim\mathcal{N}\left(0,D_{1}\right)$
is Gaussian and independent of $U_{1}$ and $Z$. Let
\[
\gamma\triangleq\frac{D_{1}\sigma_{Z}^{2}}{D_{1}\sigma_{Z}^{2}-D_{2}\left(D_{1}+\sigma_{Z}^{2}\right)},
\]
and consider an auxiliary Gaussian random variable $U'$ defined as
$U'=X'+Z_{4}$, where $Z_{4}\sim\mathcal{N}\left(0,\gamma D_{2}\right)$.
Moreover, we define $Y'\triangleq Y-U_{1}=X'+Z$ and $Y'\sim\mathcal{N}\left(0,D_{1}+\sigma_{Z}^{2}\right)$.
Then we can apply the minimum mean square error (MMSE) rescaling parameter
$\alpha=\frac{D_{1}}{D_{1}+\sigma_{Z}^{2}}$ to $Y'$. As a result,
we obtain $X'=\alpha Y'+Z_{3}$, where $Z_{3}\sim\mathcal{N}\left(0,\alpha\sigma_{Z}^{2}\right)$.
We can also write $U'=\alpha Y'+Z_{5}$, where $Z_{5}\sim\mathcal{N}\left(0,\gamma D_{2}+\alpha\sigma_{Z}^{2}\right)$,
which requires an AWGN capacity-achieving code from $\alpha Y'$ to
$U'$. This test channel is depicted in Fig. \ref{fig:GaussianTestChannel_1}. 

The final reconstruction at Decoder 2 is given by $\hat{X}_{2}=U_{1}+\alpha Y'+\frac{1}{\gamma}\left(U'-\alpha Y'\right).$
Note that $\frac{1}{\gamma}\left(U'-\alpha Y'\right)$ is a scaled
version of $Z_{5}$, which is independent of $Y'$. Thus, the variance
of $\alpha Y'+\frac{1}{\gamma}\left(U'-\alpha Y'\right)$ is $\alpha D_{1}+\frac{1}{\gamma^{2}}\left(\gamma D_{2}+\alpha\sigma_{Z}^{2}\right)=D_{1}-D_{2}$.
Therefore, we have $X-U_{1}=\hat{X}_{2}-U_{1}+\mathcal{N}\left(0,D_{2}\right)$
as we desired. Furthermore, the required data rate for Decoder 2 is
then given by $R_{2}>I\left(U';X'\right)-I\left(U';\alpha Y'\right)=\frac{1}{2}\log\left(\frac{D_{1}\sigma_{Z}^{2}}{D_{2}\left(D_{1}+\sigma_{Z}^{2}\right)}\right).$ 

\begin{figure}
\begin{centering}
\includegraphics[scale=0.7]{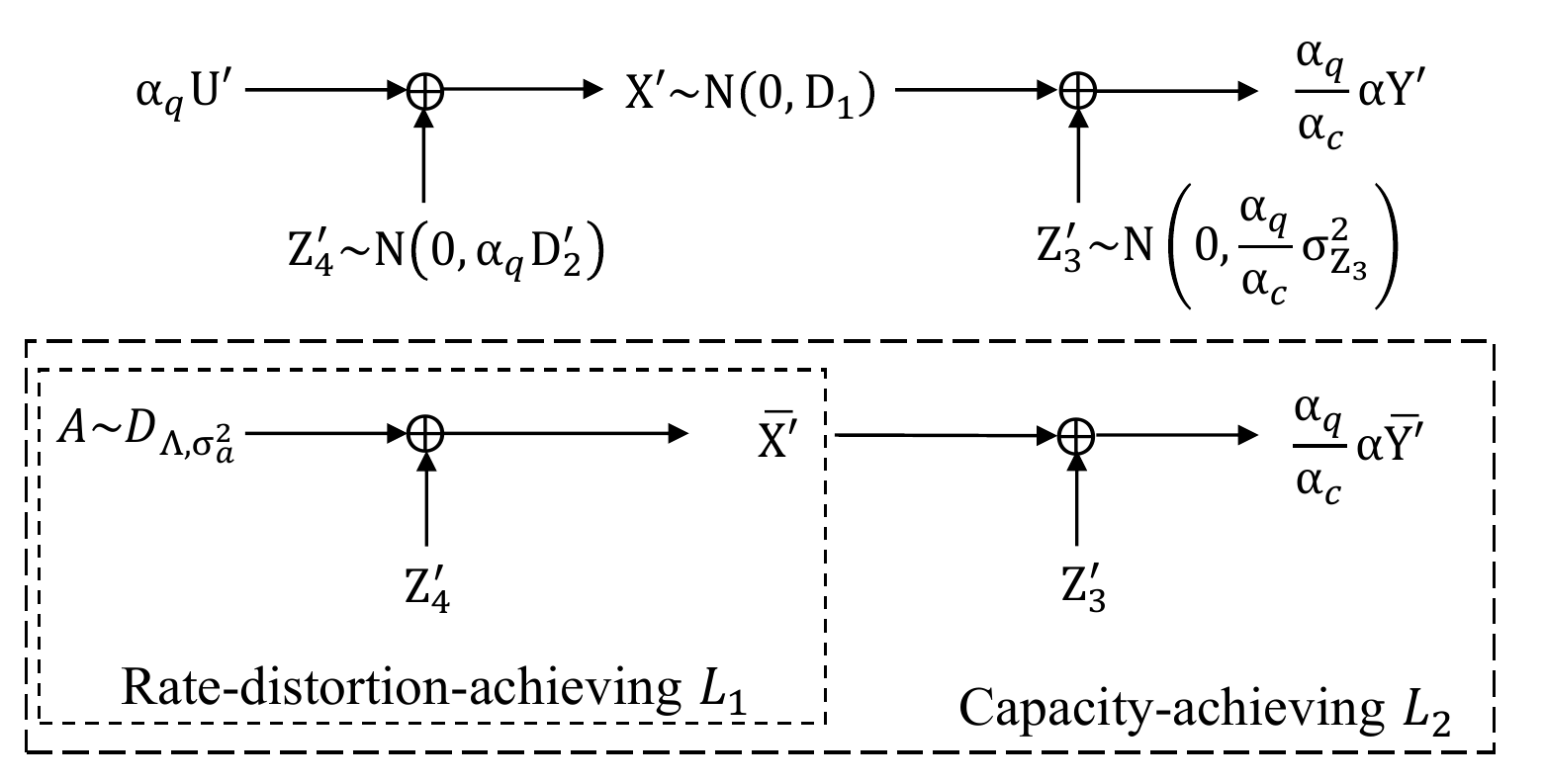}
\par\end{centering}
\caption{A reversed solution of the test channels in order to construct polar
lattices. \label{fig:GaussianTestChannel_2} }

\end{figure}

Note that $U'$ is a continuous Gaussian random variable which is
impractical for the design of polar lattices. Hence, we use the discrete
Gaussian distribution $D_{\Lambda,\sigma_{U'}^{2}}$ to replace it.
Before that, we need to perform MMSE rescaling on $U'$ for test channels
$X'\rightarrow U'$ and $Y'\rightarrow U'$ with scales $\alpha_{q}$
and $\alpha_{c}$, respectively. Consequently, a reversed version
of the test channel in Fig. \ref{fig:GaussianTestChannel_1} can be
derived, as depicted in Fig. \ref{fig:GaussianTestChannel_2}, where
\[
\alpha_{q}=\frac{D_{1}}{D_{1}+D'_{2}}=\frac{D_{1}\left(\sigma_{Z}^{2}-D_{2}\right)-D_{2}\sigma_{Z}^{2}}{D_{1}\left(\sigma_{Z}^{2}-D_{2}\right)}
\]
and 
\[
\alpha_{c}=\frac{D_{1}^{2}}{\left(D_{1}+D'_{2}\right)\left(D_{1}+\sigma_{Z}^{2}\right)}=\frac{D_{1}\left(\sigma_{Z}^{2}-D_{2}\right)-D_{2}\sigma_{Z}^{2}}{\left(D_{1}+\sigma_{Z}^{2}\right)\left(\sigma_{Z}^{2}-D_{2}\right)}.
\]

The reconstruction of $X$ at Decoder 2 is as given in the following
proposition.
\begin{prop}
\label{prop:reconstruction}If we use the reversed test channel shown
in Fig \ref{fig:GaussianTestChannel_2}, the reconstruction of $X$
at Decoder 2 is given by 
\[
\hat{X}_{2}=U_{1}+\alpha_{q}U'+\eta\left(\frac{\alpha_{q}}{\alpha_{c}}\alpha Y'-\alpha_{q}U'\right),\textrm{ }\eta=\frac{D_{2}}{\sigma_{Z}^{2}}.
\]
\end{prop}
\begin{IEEEproof}
It suffices to prove that $X-U_{1}=\hat{X}_{2}-U_{1}+\mathcal{N}\left(0,D_{2}\right)$.
Since we have $X'=\alpha_{q}U'+\mathcal{N}\left(0,\alpha_{q}D'_{2}\right)$,
as illustrated in Fig \ref{fig:GaussianTestChannel_2}, showing $\hat{X}_{2}-U_{1}=\alpha_{q}U'+\mathcal{N}\left(0,\alpha_{q}D'_{2}-D_{2}\right)$
would complete the proof. We can see from Fig \ref{fig:GaussianTestChannel_2}
that $\frac{\alpha_{q}}{\alpha_{c}}\alpha Y'-\alpha_{q}U'$ is Gaussian
distributed with zero mean and variance $\alpha_{q}D'_{2}+\frac{\alpha_{q}}{\alpha_{c}}\sigma_{Z_{3}}^{2}$,
and it is independent of $U'$. We also have 
\[
\begin{aligned}\alpha_{q}D'_{2}-D_{2} & =\frac{D_{1}\left(\sigma_{Z}^{2}-D_{2}\right)-D_{2}\sigma_{Z}^{2}}{D_{1}\left(\sigma_{Z}^{2}-D_{2}\right)}\frac{D_{1}D_{2}\sigma_{Z}^{2}}{D_{1}\sigma_{Z}^{2}-D_{2}\left(D_{1}+\sigma_{Z}^{2}\right)}-D_{2}\\
 & =\frac{D_{2}^{2}}{\sigma_{Z}^{2}-D_{2}},
\end{aligned}
\]
and 
\[
\begin{aligned}\eta^{2}\left(\alpha_{q}D'_{2}+\frac{\alpha_{q}}{\alpha_{c}}\sigma_{Z_{3}}^{2}\right) & =\left(\frac{D_{2}}{\sigma_{Z}^{2}}\right)^{2}\left(\frac{D_{2}\sigma_{Z}^{2}}{\left(\sigma_{Z}^{2}-D_{2}\right)}+\frac{D_{1}+\sigma_{Z}^{2}}{D_{1}}\frac{D_{1}\sigma_{Z}^{2}}{D_{1}+\sigma_{Z}^{2}}\right)\\
 & =\frac{D_{2}^{2}}{\sigma_{Z}^{2}-D_{2}},
\end{aligned}
\]
as we desired.
\end{IEEEproof}

Based on the reversed test channel, we can replace the continuous
Gaussian random variable $\alpha_{q}U'$ with a discrete Gaussian
distributed variable $A\sim D_{\varLambda,\sigma_{a}^{2}}$, where
$\sigma_{a}^{2}=\alpha_{q}^{2}\sigma_{U'}^{2}$. Let $\bar{X'}\triangleq A+\mathcal{N}\left(0,\alpha_{q}D'_{2}\right)$
and $\frac{\alpha_{q}}{\alpha_{c}}\alpha\bar{Y'}=\bar{X'}+\mathcal{N}\left(0,\frac{\alpha_{q}}{\alpha_{c}}\sigma_{Z_{3}}^{2}\right)$.
We also define $\bar{B}\triangleq\frac{\alpha_{q}}{\alpha_{c}}\alpha\bar{Y'}$,
whose variance is $\sigma_{b}^{2}=\frac{\alpha_{q}^{2}}{\alpha_{c}^{2}}\alpha^{2}\sigma_{Y'}^{2}$.
By \cite[Lemma 1]{PolarlatticeQZ}, the distributions of $\bar{X'}$
and $\bar{Y'}$ can be made arbitrarily close to the distributions
of $X'$ and $Y'$, respectively. Therefore, polar lattices can be
designed for the source $\bar{X'}$ and the side information $\bar{Y'}$
at Decoder 2. Specifically, a rate-distortion bound-achieving polar
lattice $L_{1}$ is constructed for the source $\bar{X'}$ with distortion
$\alpha_{q}D'_{2}$, and an AWGN capacity-achieving polar lattice
$L_{2}$ is constructed for the channel $A\rightarrow\frac{\alpha_{q}}{\alpha_{c}}\alpha\bar{Y'}$,
as shown in Fig. \ref{fig:GaussianTestChannel_2}. In the end, the
reconstruction of Decoder 2 is $\check{X_{2}}=U_{1}+A+\eta\left(\bar{B}-A\right)$.
Even though the quantization noise of $L_{1}$ is not an exact Gaussian
distribution, it is shown in \cite[Theorem 4]{PolarlatticeQZ} that
the two distributions can be arbitrarily close when $N$ is sufficiently
large. Therefore, $\bar{B}-A$ can be treated as Gaussian noise independent
of $A$. By Proposition \ref{prop:reconstruction}, $\eta$ scales
$\bar{B}-A$ to $\mathcal{N}\left(0,\alpha_{q}D'_{2}-D_{2}\right)$
and by \cite[Lemma 1]{PolarlatticeQZ}, the distributions of $\check{X_{2}}$
and $\hat{X}_{2}$ can be arbitrarily close, which gives an average
distortion close to $D_{2}$. 

According to \cite[Lemma 10]{polarlatticeJ}, $L_{1}$ and $L_{2}$
can be equivalently constructed for the MMSE-rescaled channel with
Gaussian noise variances 
\[
\tilde{\sigma_{q}^{2}}=\frac{\sigma_{a}^{2}}{D_{1}}\alpha_{q}D'_{2}=\frac{\sigma_{a}^{2}\sigma_{Z}^{2}D_{2}}{D_{1}\left(\sigma_{Z}^{2}-D_{2}\right)},
\]
and 
\[
\tilde{\sigma_{c}^{2}}=\frac{\sigma_{a}^{2}}{\sigma_{b}^{2}}\left(\alpha_{q}D'_{2}+\frac{\alpha_{q}}{\alpha_{c}}\sigma_{Z_{3}}^{2}\right)=\frac{\sigma_{a}^{2}\sigma_{Z}^{4}}{\left(D_{1}+\sigma_{Z}^{2}\right)\left(\sigma_{Z}^{2}-D_{2}\right)}.
\]

The coding strategy for $L_{1}$ and $L_{2}$ can be adapted from
\cite[Section V]{PolarlatticeQZ}. We briefly describe it for completeness.
First, choose a good constellation $D_{\varLambda,\sigma_{a}^{2}}$
such that the flatness factor $\epsilon_{\Lambda}(\tilde{\sigma_{q}^{2}})$
is negligible. Let $\Lambda/\Lambda_{1}/\cdots/\Lambda_{r-1}/\Lambda_{r}/\cdots$
denote a one-dimensional binary partition chain labeled by bits $A_{1}/A_{2}/\cdots/A_{r-1}/A_{r}/\cdots$.
Therefore, $P_{A_{1:r}}$ and $A_{1:r}$ approaches $D_{\varLambda,\sigma_{a}^{2}}$
and $A$, respectively, as $r\rightarrow\infty$. Consider $N$ i.i.d
copies of $A$, and let $K_{l}^{1:N}\triangleq A_{l}^{1:N}G_{N}$
for each level $1\leq l\leq r$. For $0<\beta<0.5$, the frozen set
$\mathcal{F}_{l}^{Q}\left(\mathcal{F}_{l}^{C}\right)$, information
set $\mathcal{I}_{l}^{Q}\left(\mathcal{I}_{l}^{C}\right)$, and shaping
set $\mathcal{S}_{l}^{Q}\left(\mathcal{S}_{l}^{C}\right)$ for $L_{1}\left(L_{2}\right)$
at level $l$ can be adapted from \cite[Equation (35)]{PolarlatticeQZ}
and \cite[Equation (36)]{PolarlatticeQZ}, respectively, by replacing
$\bar{X}$ with $\bar{X'}$. 

Furthermore, according to \cite[Lemma 2]{PolarlatticeQZ}, $L_{2}$
is nested within $L_{1}$, i.e., $L_{2}\subseteq L_{1}$. By the fact
$\tilde{\sigma_{q}^{2}}\leq\tilde{\sigma_{c}^{2}}$ and \cite[Lemma 3]{polarlatticeJ},
the partition channel $\Lambda_{l-1}/\Lambda_{l}$ with noise variance
$\tilde{\sigma_{c}^{2}}$ is degraded with respect to the one with
noise variance $\tilde{\sigma_{q}^{2}}$. Therefore, we have $\mathcal{F}_{l}^{Q}\subseteq\mathcal{F}_{l}^{C}$,
$\mathcal{I}_{l}^{C}\subseteq\mathcal{I}_{l}^{Q}$, and by the definition
of shaping set, we observe that $\mathcal{S}_{l}^{Q}\subseteq\mathcal{S}_{l}^{C}$. 

The encoder can recover the auxiliary codeword $U_{1}^{1:N}$ for
Decoder 1, and obtains the realizations $x'^{1:N}$ $\left(y'^{1:N}\right)$
of $X'^{1:N}=X^{1:N}-U_{1}^{1:N}\left(Y'^{1:N}=Y^{1:N}-U_{1}^{1:N}\right)$
from given realizations of variables $X^{1:N}\left(Y^{1:N}\right)$,
respectively. The encoder recovers $k_{l}^{1:N}$ from $l=[r]$ successively
according to the random rounding quantization rules given in \cite[Equations (13), (14), (17) and (18)]{PolarlatticeQZ}.
Note that $x'^{1:N}$ as realization of $\bar{X'}{}^{1:N}$ is acceptable
since the distributions of $X'$ and $\bar{X'}$ are arbitrarily close.
Also, according to \cite[Theorem 2]{PolarlatticeQZ}, replacing the
random rounding rule with MAP decision to obtain $k_{l}^{\mathcal{S}_{l}^{Q}}$
will not affect \cite[Theorem 5]{polarlatticeJ} and \cite[Theorem 6]{polarlatticeJ}.
Consequently, the coding scheme for Decoder 2 for the Gaussian Heegard-Berger
problem can be summarized as following:

\textit{\uline{Encoding:}} From the $N$-dimensional i.i.d. source
vector $X^{1:N}$, the encoder recovers the auxiliary codeword $U_{1}^{1:N}$
employing a quantization polar lattice for source $X$ with variance
$\sigma_{X}^{2}$ and distortion $D_{1}$, and obtains $X'^{1:N}$
and $Y'^{1:N}$. Next, the encoder evaluates $K_{l}^{\mathcal{I}_{l}^{Q}}$
by random rounding and sends $K_{l}^{\mathcal{I}_{l}^{Q}\backslash\mathcal{I}_{l}^{C}}$
to the decoders. 

\textit{\uline{Decoding:}} By the pre-shared $K_{l}^{\mathcal{F}_{l}^{Q}}$
and received $K_{l}^{\mathcal{I}_{l}^{Q}\backslash\mathcal{I}_{l}^{C}}$,
Decoder 2 recovers $K_{l}^{\mathcal{I}_{l}^{C}}$ and $K_{l}^{\mathcal{S}_{l}^{Q}}$
from the side information $\bar{B}^{1:N}$ with vanishing error probability,
by using SC decoding for Gaussian channels \cite{polarlatticeJ}.
At each level, Decoder 2 obtains $K_{l}^{1:N}$, and $A^{1:N}$ can
be recovered according to \cite[Equation (38)]{PolarlatticeQZ}. Finally,
the reconstruction of Decoder 2 is 
\begin{equation}
\check{X_{2}}^{1:N}=U_{1}^{1:N}+A^{1:N}+\eta\left(\bar{B}^{1:N}-A^{1:N}\right).\label{eq:GaussianReconstruction}
\end{equation}

According to \cite[Lemma 8]{polarlatticeJ}, the encoding and decoding
complexities of polar lattices remain to be $O\left(N\log N\right)$.

As for the transmission rate of this scheme, the rate $R_{1}$ for
Decoder 1 can be arbitrarily close to $\frac{1}{2}\log\left(\frac{\sigma_{X}^{2}}{D_{1}}\right)$
according to \cite[Theorem 4]{PolarlatticeQZ}. By the same argument,
the rate $R_{L_{1}}$ of $L_{1}$ can be arbitrarily close to $\frac{1}{2}\log\left(\frac{D_{1}}{\alpha_{q}D'_{2}}\right)$
when the flatness factor is negligible. By \cite[Theorem 7]{polarlatticeJ},
the rate $R_{L_{2}}$ of the capacity-achieving lattice $L_{2}$ can
be arbitrarily close to $\frac{1}{2}\log\left(\frac{\sigma_{b}^{2}}{\alpha_{q}D'_{2}+\frac{\alpha_{q}}{\alpha_{c}}\sigma_{Z_{3}}^{2}}\right)$
with a negligible flatness factor. Since $L_{2}\subseteq L_{1}$,
the rate for Decoder 2 after some tedious calculations is given by
\[
\begin{aligned}R_{2}=R_{L_{1}}-R_{L_{2}} & \rightarrow\frac{1}{2}\log\left(\frac{D_{1}\left(\sigma_{Z}^{2}-D_{2}\right)}{D_{2}\sigma_{Z}^{2}}\right)-\frac{1}{2}\log\left(\frac{\left(D_{1}+\sigma_{Z}^{2}\right)\left(\sigma_{Z}^{2}-D_{2}\right)}{\sigma_{Z}^{4}}\right)\\
 & \rightarrow\frac{1}{2}\log\left(\frac{D_{1}\sigma_{Z}^{2}}{D_{2}\left(D_{1}+\sigma_{Z}^{2}\right)}\right),
\end{aligned}
\]
and the total rate for the Gaussian Heegard-Berger problem is 
\[
R_{1}+R_{2}\rightarrow\frac{1}{2}\log\left(\frac{\sigma_{X}^{2}\sigma_{Z}^{2}}{D_{2}\left(D_{1}+\sigma_{Z}^{2}\right)}\right),
\]
which is the same as (\ref{eq:Rhb_boundGaussian}). 

Next, we give the main theorem of the Gaussian Heegard-Berger problem
for the non-degenerate region. 
\begin{thm}
Let $\left(X,Y,Z,D_{1},D_{2}\right)$ be as specified in Section \ref{subsec:The-Gaussian-Case}.
For any rate $R_{1}>\frac{1}{2}\log\left(\frac{\sigma_{X}^{2}}{D_{1}}\right)$,
there exists a polar lattice code at rate $R_{1}$ with sufficiently
large blocklength, whose expected distortion is arbitrarily close
to $D_{1}$ and the number of partition levels is $O\left(\log\log N\right)$.
Let $\Lambda/\Lambda_{1}/\cdots/\Lambda_{r-1}/\Lambda_{r}$ be a one-dimensional
binary partition chain of a lattice $\Lambda$ such that $\epsilon_{\Lambda}(\tilde{\sigma_{q}^{2}})=O\left(2^{-\sqrt{N}}\right)$
and $r=O\left(\log N\right)$. For any $0<\beta'<\beta<0.5$, there
exist nested polar lattices $L_{1}$ and $L_{2}$ with a rate spread
$R_{2}=R_{L_{1}}-R_{L_{2}}$ arbitrarily close to $\frac{1}{2}\log\left(\frac{D_{1}\sigma_{Z}^{2}}{D_{2}\left(D_{1}+\sigma_{Z}^{2}\right)}\right)$
such that the expected distortion $D_{Q_{2}}$ satisfies $D_{Q_{2}}\leq D_{2}+O\left(2^{-N^{\beta'}}\right)$. 
\end{thm}
\begin{IEEEproof}
The achievability of the rate-distortion function for Decoder 1 follows
from \cite[Theorem 4]{PolarlatticeQZ}. The proof of the achievability
for Decoder 2 can be adapted from \cite[Theorem 5]{PolarlatticeQZ},
by considering the test channel depicted in Fig. \ref{fig:GaussianTestChannel_2}
and the reconstruction as given by (\ref{eq:GaussianReconstruction}).
It is worth mentioning that the requirements $\epsilon_{\Lambda}(\tilde{\sigma_{q}^{2}})=O\left(2^{-\sqrt{N}}\right)$
and $r=O\left(\log N\right)$ are given by \cite[Proposition 2]{PolarlatticeQZ}
to guarantee a sub-exponentially decaying error probability for the
lattice design for Decoder 2. 
\end{IEEEproof}

\section{Conclusion}

We presented nested polar codes and polar lattices that achieve the
rate-distortion function for the binary and Gaussian Heegard-Berger
problems, respectively. Different from the code constructions for
the Wyner-Ziv problem \cite{polarchannelandsource} and \cite{PolarlatticeQZ},
we took advantage of the reconstruction at Decoder 1 to build the
nested structure that achieves the rate-distortion function for Decoder
2. The proposed schemes achieve the HBRDF in the entire non-degenerate
regions for both DSBS and Gaussian sources. 

Finally, the Kaspi problem in \cite{kaspi1994rate} is regarded as
a generalization of the Heegard-Berger problem, where the encoder
may also have access to the side information. The explicit rate-distortion
functions for the Kaspi problem with Gaussian and binary sources have
been given in \cite{Perron2005KaspiGaussianRef} and \cite{Perron2007KaspiBinaryRef},
respectively. We will study the construction of polar codes and polar
lattices for the Kaspi problem in our future work.

\appendices{}

\section{Proof of Theorem \ref{thm:OptimalityDecoder2}}

First, we show that the distortion $D_{2}$ can be achieved. Since
$U_{b}^{1:N}=W^{1:N}G_{N}$ gives a one-to-one mapping between $W^{1:N}$
and $U_{b}^{1:N}$, expression (\ref{eq:L2_distance}) is equivalent
to 
\begin{equation}
\mathbb{V}\left(P_{U_{a}^{1:N},U_{b}^{1:N},X^{1:N},U_{1}^{1:N}},Q_{U_{a}^{1:N},U_{b}^{1:N},X^{1:N},U_{1}^{1:N}}\right)=O\left(2^{-N^{\beta'}}\right).\label{eq:L2_distanceV2}
\end{equation}
From the coding scheme presented in Section \ref{subsec:Proposed-Coding-Scheme-Region I},
we assume that $K_{\mathcal{I}_{C_{a}}}$ and $W_{\mathcal{I}_{C_{b}}}$can
be correctly decoded by using side information, and $K_{\mathcal{S}_{S_{a}}}$
and $W_{\mathcal{S}_{S_{b}}}$ can be recovered by the MAP rule. Therefore,
Decoder 2 can recover $U_{a}^{1:N}$ and $U_{b}^{1:N}$ with the joint
distribution $Q_{U_{a}^{1:N},U_{b}^{1:N},X^{1:N},U_{1}^{1:N}}$. Denote
by $Q_{U_{a}^{1:N},U_{b}^{1:N},X^{1:N},U_{1}^{1:N},Y^{1:N}}$ the
resulting distribution when the encoder performs compression at each
level, i.e., compresses $X^{1:N}$ to $W_{\mathcal{I}_{S_{b}}}$.
Let $P_{U_{a}^{1:N},U_{b}^{1:N},X^{1:N},U_{1}^{1:N},Y^{1:N}}$ denote
the joint distribution when the encoder does not perform compression.
For simplicity, we denote random variables $U_{1}^{1:N}$, $U_{a}^{1:N}$
and $U_{b}^{1:N}$ by $U_{1,a,b}^{1:N}$. 

\begin{align*}
 & 2\mathbb{V}\left(P_{U_{1,a,b}^{1:N},X^{1:N},Y^{1:N}},Q_{U_{1,a,b}^{1:N},X^{1:N},Y^{1:N}}\right)\\
 & =\sum_{u_{1,a,b}^{1:N},x^{1:N},y^{1:N}}\left|P\left(u_{1,a,b}^{1:N},x^{1:N},y^{1:N}\right)-Q\left(u_{1,a,b}^{1:N},x^{1:N},y^{1:N}\right)\right|\\
 & =\sum_{u_{1,a,b}^{1:N},x^{1:N},y^{1:N}}\left|P\left(u_{1,a,b}^{1:N},x^{1:N}\right)P\left(y^{1:N}|u_{1,a,b}^{1:N},x^{1:N}\right)-Q\left(u_{1,a,b}^{1:N},x^{1:N}\right)Q\left(y^{1:N}|u_{1,a,b}^{1:N},x^{1:N}\right)\right|.
\end{align*}

According to the Markov chain $Y\leftrightarrow X\leftrightarrow U_{1,a,b}^{1:N}$,
we have 
\[
P\left(y^{1:N}|u_{1,a,b}^{1:N},x^{1:N}\right)=Q\left(y^{1:N}|u_{1,a,b}^{1:N},x^{1:N}\right)=P\left(y^{1:N}|x^{1:N}\right).
\]
Therefore, 
\begin{equation}
\mathbb{V}\left(P_{U_{1,a,b}^{1:N},X^{1:N},Y^{1:N}},Q_{U_{1,a,b}^{1:N},X^{1:N},Y^{1:N}}\right)=\mathbb{V}\left(P_{U_{1,a,b}^{1:N},X^{1:N}},Q_{U_{1,a,b}^{1:N},X^{1:N}}\right)=O\left(2^{-N^{\beta'}}\right).\label{eq:distanceWithSI}
\end{equation}

The reconstructions of two levels are $U_{a}^{1:N}$ and $U_{b}^{1:N}$
(i.e., denoted by $U_{a,b}^{1:N}$), and the average distortion $D_{P}$
achieved by $P_{U_{1,a,b}^{1:N},X^{1:N},Y^{1:N}}$ is given by 
\begin{align*}
D_{P} & =\frac{1}{N}\sum_{u_{1,a,b}^{1:N},x^{1:N},y^{1:N}}P_{U_{1,a,b}^{1:N},X^{1:N},Y^{1:N}}\left(u_{1,a,b}^{1:N},x^{1:N},y^{1:N}\right)d\left(U_{a,b}^{1:N},x^{1:N}\right)\\
 & =\frac{1}{N}\sum_{u_{a,b}^{1:N},x^{1:N}}P_{U_{a,b}^{1:N},X^{1:N}}\left(u_{a,b}^{1:N},x^{1:N}\right)d\left(U_{a,b}^{1:N},x^{1:N}\right)\\
 & =\frac{1}{N}N\sum_{u_{a,b},x}P_{U_{a,b},X}\left(u_{a,b},x\right)d\left(U_{a,b},x\right)\\
 & =D_{2}.
\end{align*}
Note that the last equality holds due to the constrains $\left(\theta-\theta_{1}\right)\alpha+\theta_{1}\mu+\left(1-\theta\right)p=D_{2}$
of Theorem \ref{thm:HBRD_region_I}. This is reasonable because $D_{P}$
is achieved when the encoder does not perform any compression. Combined
with (\ref{eq:distanceWithSI}), the expected distortion $D_{Q}$
achieved by $Q_{U_{1,a,b}^{1:N},X^{1:N},Y^{1:N}}$ satisfies 
\begin{align*}
D_{Q}-D_{P} & =\frac{1}{N}\sum_{u_{1,a,b}^{1:N},x^{1:N},y^{1:N}}\left(Q_{U_{1,a,b}^{1:N},X^{1:N},Y^{1:N}}-P_{U_{1,a,b}^{1:N},X^{1:N},Y^{1:N}}\right)d\left(U_{a,b}^{1:N},x^{1:N}\right)\\
 & \leq\frac{1}{N}Nd_{\max}\sum_{u_{1,a,b}^{1:N},x^{1:N},y^{1:N}}\left|P_{U_{1,a,b}^{1:N},X^{1:N},Y^{1:N}}-Q_{U_{1,a,b}^{1:N},X^{1:N},Y^{1:N}}\right|\\
 & =O\left(2^{-N^{\beta'}}\right).
\end{align*}

Next we show that the decoder can recover $U_{a}^{i\in\mathcal{I}_{C_{a}}}$
and $U_{b}^{i\in\mathcal{I}_{C_{b}}}$ with a sub-exponentially decaying
block error probability. 
\[
\begin{aligned}2\mathbb{V}\left(P_{U_{1,a,b}^{1:N},Y^{1:N}},Q_{U_{1,a,b}^{1:N},Y^{1:N}}\right) & =\sum_{u_{1,a,b}^{1:N},y^{1:N}}\left|P\left(u_{1,a,b}^{1:N},y^{1:N}\right)-Q\left(u_{1,a,b}^{1:N},y^{1:N}\right)\right|\\
 & =\sum_{u_{1,a,b}^{1:N},y^{1:N}}\left|\sum_{x^{1:N}}\left[P\left(u_{1,a,b}^{1:N},x^{1:N},y^{1:N}\right)-Q\left(u_{1,a,b}^{1:N},x^{1:N},y^{1:N}\right)\right]\right|\\
 & \leq\sum_{u_{1,a,b}^{1:N},y^{1:N}}\sum_{x^{1:N}}\left|P\left(u_{1,a,b}^{1:N},x^{1:N},y^{1:N}\right)-Q\left(u_{1,a,b}^{1:N},x^{1:N},y^{1:N}\right)\right|\\
 & =O\left(2^{-N^{\beta'}}\right).
\end{aligned}
\]

Let $E_{P}^{a}\left[Pe\right]$ and $E_{P}^{b}\left[Pe\right]$ denote
the expectation error probability as a result of the distribution
$P_{U_{1,a,b}^{1:N},Y^{1:N}}$ at level 1 and 2, respectively. Take
$E_{P}^{b}\left[Pe\right]$ as an example to show the decaying error
probability. Let $\mathcal{E}_{i}$ denote the set of random variables
$\left(u_{b}^{1:N},u_{1,a}^{1:N},y^{1:N}\right)$ such that the SC
decoding error occurred at the $i$th bit. Hence the block error event
is defined by $\mathcal{E}^{b}\triangleq\cup_{i\in\mathcal{I}_{C_{b}}}\mathcal{E}_{i}$,
and the expectation of decoding block error probability over all random
mapping is given by 
\[
\begin{aligned}E_{P}^{b}\left[Pe\right] & =\sum_{u_{1,a,b}^{1:N},y^{1:N}}P_{U_{1,a,b}^{1:N},Y^{1:N}}\left(u_{1,a,b}^{1:N},y^{1:N}\right)\boldsymbol{1}\left[\left(u_{b}^{1:N},u_{1,a}^{1:N},y^{1:N}\right)\in\mathcal{E}^{b}\right]\\
 & \leq\sum_{i\in\mathcal{I}_{C_{b}}\cup\mathcal{S}_{S_{b}}}\sum_{u_{1,a,b}^{1:N},y^{1:N}}P_{U_{1,a,b}^{1:N},Y^{1:N}}\left(u_{1,a,b}^{1:N},y^{1:N}\right)\boldsymbol{1}\left[\left(u_{b}^{1:N},u_{1,a}^{1:N},y^{1:N}\right)\in\mathcal{E}_{i}\right]\\
 & \leq\sum_{i\in\mathcal{I}_{C_{b}}\cup\mathcal{S}_{S_{b}}}\sum_{u_{b}^{1:i},u_{1,a}^{1:N},y^{1:N}}P\left(u_{b}^{1:i-1},u_{1,a}^{1:N},y^{1:N}\right)P\left(u_{b}^{i}|u_{b}^{1:i-1},u_{1,a}^{1:N},y^{1:N}\right)\\
 & \hspace{3cm}\cdot\boldsymbol{1}\left[P\left(u_{b}^{i}|u_{b}^{1:i-1},u_{1,a}^{1:N},y^{1:N}\right)\leq P\left(u_{b}^{i}\oplus1|u_{b}^{1:i-1},u_{1,a}^{1:N},y^{1:N}\right)\right]\\
 & \leq\sum_{i\in\mathcal{I}_{C_{b}}\cup\mathcal{S}_{S_{b}}}\sum_{u_{b}^{1:i},u_{1,a}^{1:N},y^{1:N}}P\left(u_{b}^{1:i-1},u_{1,a}^{1:N},y^{1:N}\right)P\left(u_{b}^{i}|u_{b}^{1:i-1},u_{1,a}^{1:N},y^{1:N}\right)\\
 & \hspace{3cm}\cdot\sqrt{\frac{P\left(u_{b}^{i}\oplus1|u_{b}^{1:i-1},u_{1,a}^{1:N},y^{1:N}\right)}{P\left(u_{b}^{i}|u_{b}^{1:i-1},u_{1,a}^{1:N},y^{1:N}\right)}}\\
 & \leq N\cdot Z\left(U_{b}^{i}|U_{b}^{1:i-1},U_{1,a}^{1:N},Y^{1:N}\right)\\
 & =O\left(2^{-N^{\beta'}}\right).
\end{aligned}
\]
Following the same arguments, we also have $E_{P}^{a}\left[Pe\right]=O\left(2^{-N^{\beta'}}\right)$.
Therefore, by this union bound, we obtain the two-stage decoding block
error probability $E_{P}\left[Pe\right]=O\left(2^{-N^{\beta'}}\right)$. 

Let $Pe^{HB}$ denote the expectation of error probability caused
by $Q_{U_{1,a,b}^{1:N},Y^{1:N}}$, which is an average over all choices
of $U_{a}^{i\in\mathcal{F}_{C_{a}}}$, $U_{a}^{i\in\mathcal{S}_{C_{a}}\backslash\mathcal{S}_{S_{a}}}$,
$U_{b}^{i\in\mathcal{F}_{C_{b}}}$ and $U_{b}^{i\in\mathcal{S}_{C_{b}}\backslash\mathcal{S}_{S_{b}}}$at
each level. Let $\mathcal{E}$ denote the set of random variables
$\left(u_{b}^{1:N},u_{1,a}^{1:N},y^{1:N}\right)$ such that a decoding
error occurs. Then we have

\[
\begin{aligned}Pe^{HB}-E_{P}\left[Pe\right] & =\sum_{u_{1,a,b}^{1:N},y^{1:N}}\left(Q\left(u_{1,a,b}^{1:N},y^{1:N}\right)-P\left(u_{1,a,b}^{1:N},y^{1:N}\right)\right)\cdot\boldsymbol{1}\left[\left(u_{b}^{1:N},u_{1,a}^{1:N},y^{1:N}\right)\in\mathcal{E}\right]\\
 & \leq2\mathbb{V}\left(P_{U_{1,a,b}^{1:N},Y^{1:N}},Q_{U_{1,a,b}^{1:N},Y^{1:N}}\right)\\
 & \leq O\left(2^{-N^{\beta'}}\right).
\end{aligned}
\]

As for the rates, we have $\frac{\left|\mathcal{I}_{S_{a}}\right|}{N}\xrightarrow{N\rightarrow\infty}I\left(U_{a};X,U_{1}\right)$
and $\frac{\left|\mathcal{I}_{C_{a}}\right|}{N}\xrightarrow{N\rightarrow\infty}I\left(U_{a};Y,U_{1}\right)$
at the first level. Therefore, we have 

\[
\frac{\left|\mathcal{I}_{S_{a}}\right|-\left|\mathcal{I}_{C_{a}}\right|}{N}\xrightarrow{N\rightarrow\infty}I\left(X;U_{a}|U_{1},Y\right).
\]

For the second level, $\frac{\left|\mathcal{I}_{S_{b}}\right|}{N}\xrightarrow{N\rightarrow\infty}I\left(U_{b};X,U_{1},U_{a}\right)$
and $\frac{\left|\mathcal{I}_{C_{b}}\right|}{N}\xrightarrow{N\rightarrow\infty}I\left(U_{b};Y,U_{1},U_{a}\right)$.
Thus, we have 
\[
\frac{\left|\mathcal{I}_{S_{b}}\right|-\left|\mathcal{I}_{C_{b}}\right|}{N}\xrightarrow{N\rightarrow\infty}I\left(X;U_{b}|U_{1},U_{a},Y\right).
\]

Finally, the rate of Decoder 2 is 
\[
\frac{\left|\mathcal{I}_{S_{a}}\right|-\left|\mathcal{I}_{C_{a}}\right|+\left|\mathcal{I}_{S_{b}}\right|-\left|\mathcal{I}_{C_{b}}\right|}{N}\xrightarrow{N\rightarrow\infty}I\left(X;U_{2}|U_{1},Y\right).
\]

\bibliographystyle{IEEEtran}
\bibliography{MyreffGeneral}

\end{document}